\documentclass[authoryear,review]{elsarticle}

\usepackage{natbib}
\usepackage{lineno,hyperref}
\usepackage{wrapfig}
\usepackage{amssymb,amsmath}
\usepackage{bm,upgreek}
\usepackage{mathrsfs}
\usepackage{graphics}
\usepackage{amsfonts}
\usepackage{color}
\usepackage{multirow}
\usepackage{lipsum}
\usepackage{booktabs}
\usepackage{floatpag}
\usepackage{gensymb}
\usepackage{soul}
\usepackage{pdflscape}
\usepackage{rotating}
\usepackage{tikz}
\usepackage{geometry}
\usepackage{caption}
\usepackage{pdfpages}

\soulregister\cite7
\soulregister\ref7
\soulregister\eqref7

\usepackage[symbol]{footmisc}

\usepackage{array}
\usepackage{etoolbox}

\graphicspath{{Fig/}}

\newcommand\rd{{\rm{d}}}

\DeclareMathOperator{\tr}{tr}

\definecolor{myviolet}{RGB}{255,0,255}
\definecolor{darkgreen}{rgb}{0,0.5,0}
\definecolor{shade}{RGB}{176,176,176}

\newcommand{\circled}[2][]{%
\tikz[baseline=(char.base), text=black]{%
\node[shape = circle, draw=black, inner sep = 0.5pt]
(char) {\phantom{\ifblank{#1}{#2}{#1}}};%
\node at (char.center) {\makebox[0pt][c]{#2}};}}
\robustify{\circled}

\makeatletter
\def\ps@pprintTitle{
    \let\@oddhead\@empty
    \let\@evenhead\@empty
    \def\@oddfoot{
        \footnotesize
        Preprint submitted to \textit{Journal of the Mechanics of Physics of Solids}\hfill
    }
    \let\@evenfoot\@oddfoot
}
\makeatother

\begin{document}

%%%%%%%%%%Titlepage%%%%%%%%%%

\begin{frontmatter}

\title{Size-dependent transformation patterns in NiTi tubes under tension and bending: stereo digital image correlation experiments and modeling}

%Here are other suggestions for the title:
% Size-dependent transformation pattern in NiTi tubes under tension and bending: stereo-DIC experiments ande modeling
% Size-dependent transformation pattern in NiTi tubes under tension and bending: experiment ande modeling
% Size-dependent transformation pattern and front morphologies of NiTi tubes under tension and bending---stereo-DIC experiments and numerical simulations

\author[PAS,Ruhr]{Aslan Ahadi\corref{cor1}}
\ead{aslan.ahadi@piais.ir,aslan.ahadi@rub.de}

\author[PAS]{Elham Sarvari}
\ead{elham.sarvari@piais.ir}

\author[Ruhr]{Jan Frenzel}
\ead{jan.a.frenzel@rub.de}

\author[Ruhr]{Gunther Eggeler}
\ead{gunther.eggeler@rub.de}

\author[IPPT]{Stanis{\l}aw Stupkiewicz}
\ead{sstupkie@ippt.pan.pl}

\author[IPPT]{Mohsen Rezaee-Hajidehi\corref{cor2}}
\ead{mrezaee@ippt.pan.pl}

\cortext[cor1]{Corresponding author of the experimental part.}

\cortext[cor2]{Corresponding author of the modeling part.}

\address[PAS]{Pasargad Institute for Advanced Innovative Solutions (PIAIS),\\
Tehran, Iran.}

\address[Ruhr]{Institute for Materials, Ruhr University Bochum,\\
Universitätsstr. 150, 44801 Bochum, Germany}

\address[IPPT]{Institute of Fundamental Technological Research (IPPT), Polish Academy of Sciences,\\
Pawi\'nskiego 5B, 02-106 Warsaw, Poland.}

\begin{abstract}
The dependence of transformation pattern in superelastic NiTi tubes on tube outer diameter~$D$ and wall-thickness~$t$ is investigated through quasi-static uniaxial tension and large-rotation bending experiments. The evolution of outer-surface strain fields is synchronized with global stress--strain and moment--curvature responses using a multi-magnification, high-resolution stereo digital image correlation system at $0.5$--$2\times$ magnifications. The transformation patterns exhibit systematic size-dependent behaviors. Under tension and for a specific $D$, as the diameter-to-thickness ratio $D/t$ decreases, a decreasing number of fat/diffuse helical bands emerge, in contrast to sharp/slim bands in thin tubes. Consequently, the austenite--martensite front morphology transitions from finely-fingered to coarsely-fingered with decreasing $D/t$. Below a characteristic $D/t$, front morphology no longer exhibits patterning and phase transformation proceeds via propagation of a finger-less front. Moreover, the transformation pattern exhibits an interrelation between $D$ and $D/t$, where a front possessing diffuse fingers is observed in a thin but small tube. Under bending, both the global moment--curvature response and transformation pattern exhibit $D$- and $D/t$-dependence. {While wedge-like martensite domains consistently form across all tube sizes, their growth is noticeably limited in smaller and thicker tubes due to geometrical constraints.} A gradient-enhanced model of superelasticity is employed to analyze the distinct transformation patterns observed in tubes of various dimensions. The size-dependent behavior is explained based on {the competition between bulk and interfacial energies, and the energetic cost of accommodating martensite fingers. By leveraging an axisymmetric tube configuration as a reference energy state, the extra energy associated with the formation of fingers is quantified.}  
\end{abstract}

\begin{keyword}
Shape memory alloys; Martensitic phase transformation; Digital image correlation; Front morphology; Size effects; Finite-element analysis
\end{keyword}

\end{frontmatter}

%\linenumbers

\section{Introduction}
Superelasticity refers to the ability of shape memory alloys (SMAs) to recover large strains under applied stress~\citep{otsuka2005physical, Grunebohm2023}. This fascinating property relies on the reversible stress-induced martensitic phase transformation (PT) that proceeds through nucleation/propagation of a macroscopic martensite domain during loading of a sample, followed by its shrinkage/vanishing during unloading \citep{miyazaki1981luders,shaw1995thermomechanical,shaw1997initiation,shaw1997nucleation,brinson2004stress,li2023mechanical}. The nucleation of the martensite domain occurs under a non-equilibrium setting \citep{cross1993pattern} and is accompanied by instabilities such as a stress drop in the stress--strain response \citep{shaw1995thermomechanical,sittner2005origin}, strain localization on the surface of the specimen \citep{daly2007stress,zhang2010experimental,LI2023144418,LAPLANCHE2017143}, and temperature gradients due to the localized heat release \citep{shaw1997nucleation,ahadi2014effects} within the martensite domain. In view of the spatiotemporal heterogeneity of the thermomechanical fields, many challenges arise in integrating superelastic SMA components into industrial/medical/engineering applications; examples include jump phenomena in vibration/seismic control devices \citep{xia2015jump,xia2017thermomechanical}, poor fatigue response \citep{xia2015jump,xia2017thermomechanical,zhang2018fatigue}, and localized heating/cooling in elastocaloric cooling applications \citep{Ahadi2019}. An in-depth understanding of this complex phenomenon is essential for the reliable use of SMA components in many different fields \citep{li2023mechanical}.

The pattern of the transforming domain under stress-induced PT is governed by several factors; namely loading rate \citep{zhang2010experimental,he2010rate}, ambient temperature \citep{xiao2017situ}, microstructure \citep{ahadi2014effects,ahadi2013stress,sun2014effects}, deformation mode \citep{feng2006experimental,reedlunn2014tension,reedlunn2020axial}, and shape of the SMA component (tube, wire, strip, rod, etc.) \citep{Sedlak_20213D}. Under a quasi-static tensile load (isothermal condition), a single martensite domain nucleates and propagates \citep{shaw1997nucleation}. With increasing the strain rate, many crossing domains of martensite form \citep{zhang2010experimental,sun2012recent,tsimpoukis2024rate}, and under very high strain rates ($\dot{\varepsilon}>1$~s$^{-1}$), the formation of a homogeneous domain morphology on the specimen's surface has been reported \citep{zhang2010experimental,sun2012recent}. With increasing the ambient temperature, the domain front becomes gradually diffuse, and at sufficiently high temperatures the stress-induced PT proceeds homogeneously in the specimen \citep{bian2019comparative,bian2018digital}. It was also shown that the domain front becomes gradually diffuse with the grain size reduction to nanoscale \citep{ahadi2013stress,ahadi2014effects,sun2014effects,ahadi2015stress}. Moreover, lack of strain localization in compression and wedge-type transformation patterning under bending have been widely observed \citep{reedlunn2014tension,jiang2016effects}. 

Helical martensite domains that grow spirally are one of the most unique patterns that form in NiTi tubes under uniaxial tension \citep{feng2006experimental,li2002initiation,ng2006stress}. Following the unveiling of helical domain patterns, extensive experimental \citep{feng2006experimental,reedlunn2014tension,reedlunn2020axial,reedlunn2013superelastic,watkins2018uniaxial,bechle2014localization,bechle2016evolution,kazinakis2021buckling,jiang2017bending} and theoretical/numerical \citep{zhou2010energetics,damanpack2018snap,xiao2020rate,rezaee2023predicting} studies have been published to shed lights on the localized deformation behavior and transformation evolution in NiTi tubes. The surge in interest arises from the growing and developing applications that NiTi tubes are facilitating. Besides the conventional cardio-vascular stents and guide wires, NiTi tubes have entered energy-conversion sector as solid-state coolers with enhanced efficiencies owing to a large surface/volume ratio \citep{Cheng2023A, zhang2023highly,dall2024development,zhu2021modelling}. Recent advancements in fabrication and processing technologies have enabled the production of thin tubes with outer diameters ($D$) as small as 0.18~mm and wall thicknesses ($t$) of 0.05~mm, thereby incorporating NiTi tubes into neurosurgical procedures. In the aforementioned applications, NiTi tubes of various sizes undergo distinct modes of deformation, resulting in complex stress states \citep{reedlunn2014tension,reedlunn2020axial}. Hence, there remains a pressing need for systematic research focusing on the size-dependent deformation characteristics of NiTi tubes.

The phenomenon of size-dependent transformation patterning in superelastic NiTi tubes has been previously noted in a series of experimental studies by the group of Q.~P.~Sun \citep{li2002initiation,feng2006experimental,ng2006stress}. Significant efforts have been devoted to understanding the helical--cylindrical transitioning of the martensite domain in tubes of different wall thicknesses subjected to quasi-static tension. The scattered experimental observations have been later complemented by theoretical/numerical investigations \citep{he2009scaling,he2009effects,zhou2010energetics,DONG20121063}. Notably, valuable insights have been provided on the roles of geometrical and material (gradient-energy) parameters in governing the interplay between the bulk and gradient energies, and thereby, in shaping the equilibrium martensite domain. Despite these fundamental contributions, a systematic study that thoroughly elucidates the relation between the geometry and transformation behavior of NiTi tubes is still lacking. 

Motivated by this gap, the present study investigates the mechanical responses and transformation patterns of superelastic NiTi tubes with diameters $0.43~\text{mm} \leq D \leq 3~\text{mm}$ and diameter-to-thickness ratios $4.78 \leq D/t \leq 30$ under tension and large-rotation bending in quasi-static loading conditions. The evolution of localized strain fields associated with martensite domain nucleation/propagation is examined using an advanced multi-magnification stereo-DIC imaging system. The stereo-DIC system is based on parfocal zoom lens optics that enables unprecedented high-resolution observation of the localized strain fields along with a precise measurement of global mechanical responses across a broad range of tube sizes. It is worth noting that, to date, the DIC technique has been widely utilized to study the transformation behavior of NiTi tubular and flat geometries \citep{bechle2014localization,bechle2016evolution,reedlunn2013superelastic,reedlunn2020axial,SHARIAT2022142774}, including the investigation into the fatigue response of NiTi dog-bones of different thicknesses \citep{zhang2018fatigue}. However, to the best of our knowledge, the present study marks the first application of a stereo-DIC system to NiTi tubes spanning a wide range of diameters and thicknesses.

We employ a state-of-the-art gradient-enhanced model of superelasticity to complement and substantiate the experimental findings. The model, in its various versions, has been successfully applied to reproduce several key experimental results. These include the transformation patterns in NiTi specimens subjected to isothermal and non-isothermal tension \citep{RezaeeStupkiewicz2018Gradient,Rezaee2020Gradient}, the complex transformation evolution in NiTi tubes under combined tension--torsion \citep{rezaee2021modelling,rezaee2023predicting} and bending \citep{rezaee2024modeling}, as well as the subtle morphological features such as front striations \citep{rezaee2024modeling}. Hence the reliability of its predictions is well-established. In the present study, the model is leveraged to provide an energetic interpretation of the transformation pattern variation in tubes of different dimensions. In particular, the size- and thickness-dependent front morphology transitioning under tension is explained by comparing the energy state of the transformed tube with a hypothetical axisymmetric tube in which the formation of fingered morphology is suppressed. This novel approach provides a quantitative estimate of the energetic cost associated with the formation of intricate fingered morphology.

\section{Materials}
\subsection{Superelastic tubes}\label{sec:tubes}
Seven distinct superelastic NiTi tubes, characterized by outer diameter $D$, wall thickness $t$ and diameter-to-thickness ratio $D/t$ were acquired from VascoTube (Fast Tube List). The tubes are specified as follows, in the order $D$, $t$ and $D/t$, separated by slashes: 3.0/0.1/30, 2.5/0.13/19.23, 1.2/0.1/12, 2.5/0.28/8.93, 0.8/0.13/6.15, 1.2/0.2/6.0, and 0.43/0.09/4.78, with $D$ and $t$ given in mm, see Fig.~\ref{fig:setup}(a). As per the specifications provided by the supplier, all the tubes are manufactured in compliance with the ASTM F2063--05 standard and possess the typical slightly Ni-rich Ti49.1Ni50.9 (superelastic) composition, which has been validated through independent inductively coupled plasma (ICP) measurements.

\subsection{Differential Scanning calorimetry measurements}
The transformation temperatures of the tubes are double-checked using a Caliris 300 DSC. Approximately 20~mg from each tube was cut with a diamond saw at low speeds and the transformation temperatures are measured over the temperature range of $-140~^\circ$C to $50~^\circ$C at a cooling/heating rate of $\pm 10~^\circ$C/min. The cooling and heating thermograms of the tubes are shown in Fig.~S1 in the Supplementary Data. Upon cooling, all tubes exhibit a two-step PT from $B2$ to $R$-phase and from $R$-phase to $B19'$ at lower temperatures, as typically observed in recrystallization-annealed NiTi. All tubes have austenite finish temperature A$_\text{f}$ below room temperature (RT) that confirms the room-temperature superelasticity response of the tubes. The 2.5/0.13/19.23 and 0.43/0.09/4.78 tubes exhibit the highest and the lowest A$_\text{f}$ of $20~^{\circ}$C and $-17.1~^{\circ}$C, respectively.

\subsection{Grain size}
Grain size represents a critical microstructural characteristic that influences various intrinsic properties of superelasticity, including stress-plateau, stress hysteresis, and transformation pattern \citep{ahadi2013stress,ahadi2014effects,ahadi2015stress,sun2014effects}. By analyzing the DSC thermograms, which indicate a two-step PT behavior during cooling, alongside the stress--strain responses illustrated in Fig.~\ref{fig:mechsize} (showing plateau-type superelasticity with stress hysteresis values between 190~MPa and 250~MPa), it can be inferred that all tubes possess nanostructured microstructures with average grain sizes ranging from 20~nm to 50~nm \citep{Delville2010, Delville2011}. To confirm that the differences in average grain sizes are not significant, transmission electron microscopy (TEM) examinations are conducted using an FEI Tecnai $G^2 F20$ SuperTwin TEM. The cross-sectional surfaces of the 3.0/0.1/30, 2.5/0.13/19.23, 1.2/0.1/12, and 0.8/0.13/6.15 tubes were polished, and longitudinal thin foil specimens were extracted along the drawing direction utilizing a THERMOFISHER Dualbeam FIB-SEM SCIOS2. Subsequently, the grain size distributions were assessed from sets of dark-field images. The four tubes examined exhibit microstructures comparable to those reported by \citet{Robertson2005}, with average grain sizes ranging from 28~nm to 39~nm (see Fig.~S2 in Supplementary Data).

\section{Experimental details}\label{sec:exp}
\subsection{Speckle patterns for stereo-DIC measurements}
\begin{figure}[t]
    \vspace{0pt}
    \includegraphics[width=1\linewidth]{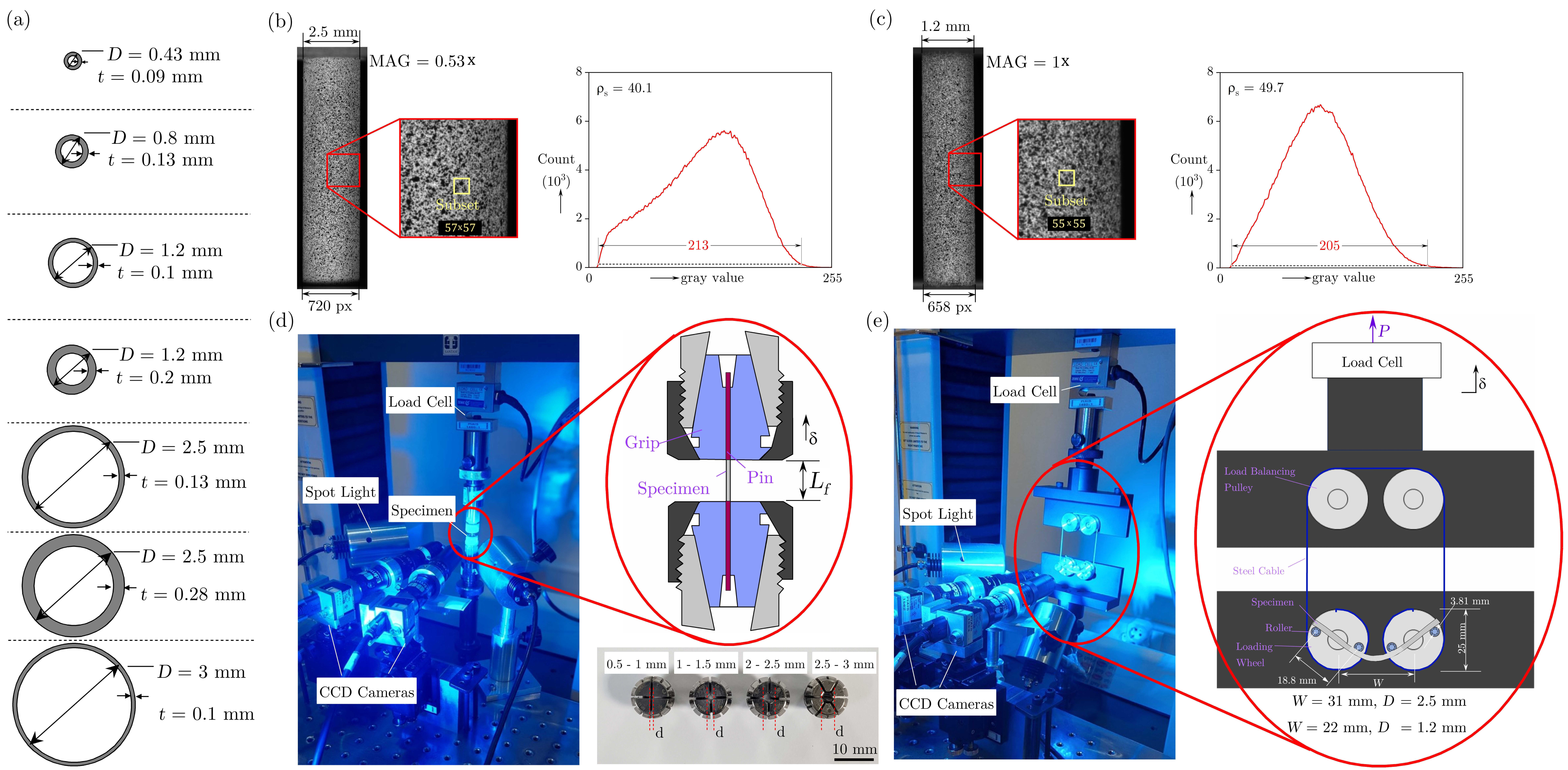}
    \caption{(a) Cross-sectional dimensions of the NiTi tubes used in the experiments. (b,c) Typical speckle patterns applied on the surface of (b) a $D = 2.5$~mm tube, imaged at $0.53\times$ magnification, and (c) a $D = 1.2$~mm tube, imaged at $1\times$ magnification. (d,e) Photographs of the experimental setups and stereo-DIC rigs for measuring mechanical responses and capturing transformation patterns under (d) quasi-static tensile setup, and (e) large-rotation bending setup.}
    \label{fig:setup}
\end{figure}

To apply speckle patterns for full-field strain measurements on the surfaces of NiTi tubes, an Iwata Custom-Micron B2 airbrush featuring a 0.18~mm paint nozzle diameter was employed. The air pressure was supplied by an Iwata IS-875 Smart Jet Pro airbrush compressor, capable of delivering pressures up to 80~psi. A thin base-coat of white paint (Golden Titanium White \#8549-4) is applied at a pressure of 40~psi with nozzle \#3 until full coverage. The black paint (Carbon Black \#2040-6) is sprayed under various spraying parameters (air pressure, nozzle number, speckling dwell time, and nozzle to specimen distance) depending on the tube dimensions (see Table~S1 in Supplementary Data). To enhance air flow, both paints are diluted with airbrush thinner (Vallejo\#71.161) in a 1/1 ratio. The quality of the speckle patterns are assessed through the mean-intensity gradient $\overline{\nabla G}$, speckle density ($\rho_\text{s}$), and gray value histogram as widely adopted by the DIC community \citep{LePage_2017Optimum,LePage_2016Cross,SarvariE2024} (see also Tables~S1 and S2). Fig.~\ref{fig:setup}(b,c) shows two typical speckle patterns along with the corresponding gray value histograms: one for a $D = 2.5$~mm tube with $\rho_\text{s}$ = 40.1\%, imaged at $0.53\times$ magnification, and the other for a $D = 1.2$~mm tube with $\rho_\text{s}$ = 49.7\%, imaged at $1\times$ magnification. Details of speckle pattern optimization for each magnification is presented in~\citet{SarvariE2024}.

\subsection{Stereo-DIC setup and post-processing}
In line with our previous study of localization in a 1~mm diameter NiTi wire, we employed a high-magnification stereo-DIC setup that can produce fields of view (FOVs) ranging from $14.9 \times 11.2$~mm$^2$ to $3.7 \times 2.81~$mm$^2$. The DIC setup utilizes parfocal zoom lens optics (VZM 200i from Edmund Optics), enabling magnification adjustments ranging from 0.5 to $2\times$ through a zoom ring, while maintaining a constant working distance of 90~mm \citep{SarvariE2024}. Images of the speckled tubes are captured with two 12 Megapixels Basler Ace CCD cameras (acA4024-29) containing $3036 \times 4024$ pixel array of square 1.85 $\mu$m pixels. The cameras are mounted on rotation stages for the stereo-angle adjustment. The rotation stages are screwed on micro-positioners for XYZ adjustments. A stereo-angle of 17$~^{\circ}$ is opted for all the stereo-DIC measurements during tension and bending experiments. Two high-intensity blue spot-lights along with a high-intensity long working distance blue light are used for illuminations, as shown in Fig.~\ref{fig:setup}. The load measurements and images are synchronized using an NI-9238 National Instrument cDAQ device. Calibration of stereo-system parameters are performed with custom-built dotted patterns tailored with respect to the FOV, as thoroughly described in \citet{SarvariE2024}. Images are captured in Pylonviewer software and data analysis and post-processing are performed in Vic-3D, which employs a cross-correlation algorithm based on the normalized sum of squared differences proposed by \citet{sutton2009image}. The strains reported here are the Biot strains along the longitudinal direction of the tubes. The magnified images in Fig.~\ref{fig:setup}(b) and (c) represent a portion of the actual images used for the DIC analysis and show typical subsets used for the correlation analysis. Details of the correlation analysis for each tube dimension are presented in Tables~S1 and S2.

\subsection{Quasi-static uniaxial tension experiments}
A Santam loading frame with a nominal capacity of 5~kN operating in a displacement-controlled mode is utilized for all experiments. The axial force $P$ is recorded with two load cells: a 500~kgf load-cell during testing of 3.0/0.1/30, 2.5/0.13/19.23, 2.5/2.8/8.93, 1.2/0.1/12, and 1.2/0.2/6.0 tubes and a 20~kgf load cell for 0.8/0.13/6.15 and 0.43/0.09/4.78 tubes. Four ER-11 steel collet chucks are used to grip the tubes with different diameters as shown in Fig.~\ref{fig:setup}(d). Dowel pins with diameters matching the inner diameter of the tubes are inserted in the gripped region to increase the stiffness of the tube walls against gripping. The free (viewable) length $L_\text{f}$ is the free length between the upper and lower grips and varies for testing tubes with different diameters $D$, as listed in Table~S1. Care was taken to avoid pre-torsion while screwing the nut. To avoid latent heat effect on the stress--strain responses and domain pattern morphologies, a sufficiently low elongation rate of $\dot \delta=0.05$~mm/min is used for all quasi-static tests. The maximum stress is set to $550$~MPa for all tubes. To ensure the reliability of the experimental results, two tests (denoted as T1 and T2) are conducted for each tube dimension, with the free length $L_\text{f}$ being nearly the same (see Table~S1).

\subsection{Bending}\label{sec:bend}
Three tubes, namely the 2.5/0.28/8.93, 1.2/0.1/12, and 1.2/0.2/6.0 tubes are tested utilizing two bending fixtures similar to that designed by \citet{reedlunn2014tension}, as shown in Fig.~\ref{fig:setup}(e). It is worth noting that in the early stages of the bending experiments on the 3.0/0.1/30 and 2.5/0.13/19.23 tubes, the tubes' wall crush at one of the contact points with the rollers, due to the insufficient rigidity of the walls. The crushing is observed repeatedly across multiple experiments. Therefore, these tubes are omitted from further testing. On the other hand, achieving adequately large curvatures in bending of 0.8/0.13/6.15 and 0.43/0.09/4.78 tubes requires fixtures with smaller loading wheels and smaller $W$ values, which are not available. Consequently, these tubes are also excluded from the study. The 2.5/0.28/8.93 tube is tested utilizing the fixture with a loading wheel diameter of $D_\text{w}=25$~mm, a distance between the centers of the loading wheels of $W=31$~mm, and a steel cable diameter of $D_\text{c} = 0.7$~mm (see Fig.~\ref{fig:setup}(e)). The 1.2/0.1/12, and 1.2/0.2/6.0 tubes are tested with another fixture: $D_\text{w}=21.5$~mm, $W=22$~mm, and $D_\text{c}=0.4$~mm. A slow cross-head displacement rate of $\dot \delta=0.1$~mm/min is used for all bending experiments. The stereo-DIC setup is the same as that utilized in quasi-static tension experiments. The moment~$M$ applied to the tube is calculated via $M=P(D_\text{w} + D_\text{c})/4$. The average curvature~$\overline{\kappa}$ is calculated via $\overline{\kappa}=(\varphi_2 - \varphi_1)/L_\text{e}$, where $\varphi_i$ are bending angles (measured via DIC analysis) and $L_\text{e}$ is the length of the area of interest (AOI) that is used in stereo-DIC calculations (see Table~S2).

\section{Results and discussion}

\subsection{Quasi-static stress--strain responses}
\begin{figure}[!t]
\centering
    \hspace*{0pt}
    \includegraphics[width=1\textwidth]{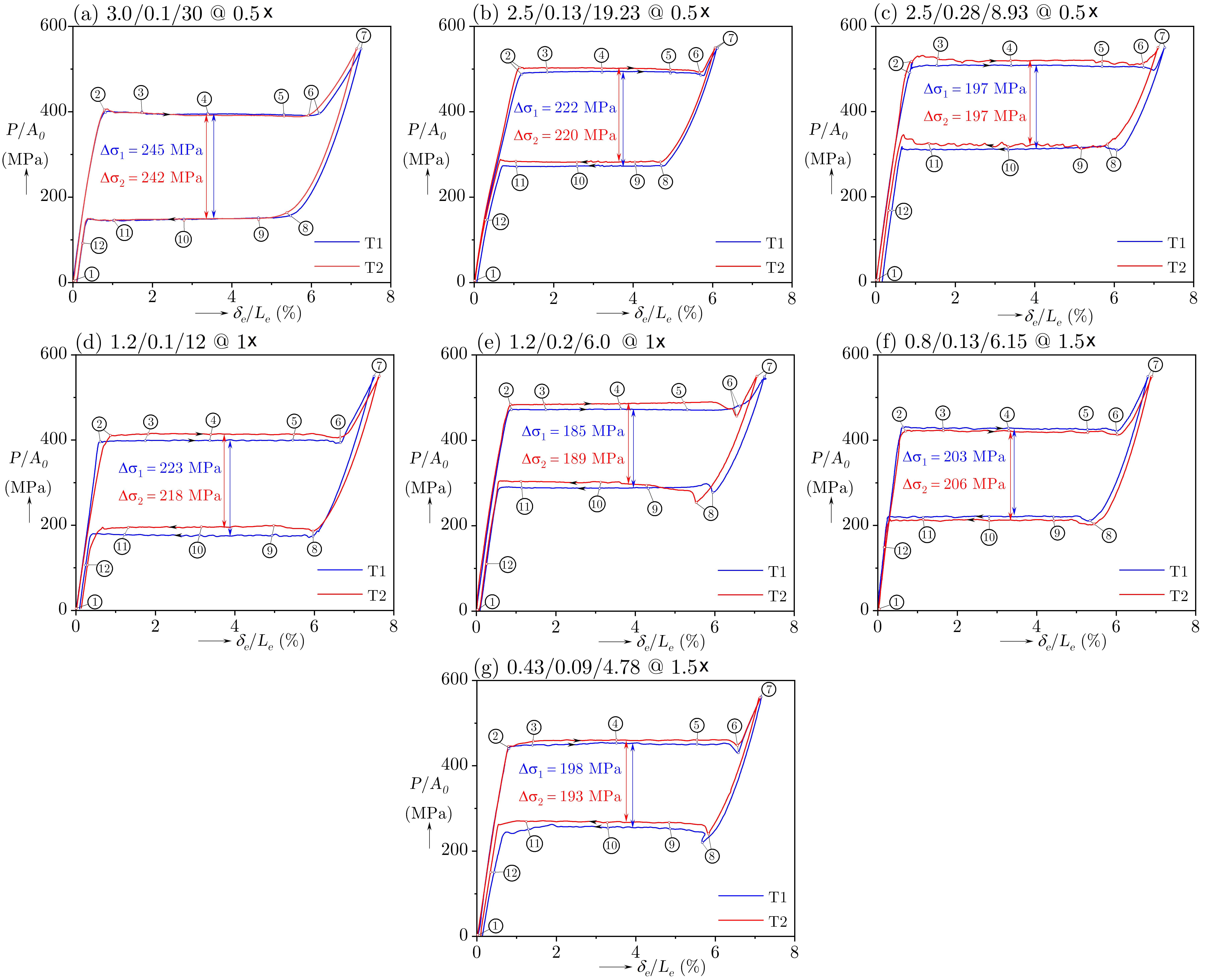}
    \caption{Quasi-static stress--strain responses of superelastic NiTi tubes with different dimensions at $T_\text{test}=22~^\circ$C. The alignment of the responses from the T1 and T2 tests, corresponding to slightly different $L_\text{e}$ (see Table~S1), demonstrate the repeatability of the experiments.}
    \label{fig:mechsize}
\end{figure}

Fig.~\ref{fig:mechsize} shows the overall stress--strain responses of the NiTi tubes with $0.43~\text{mm} \leq D \leq 3~\text{mm}$ and $4.78 \leq D/t \leq 30$ under quasi-static loading and unloading. The vertical axis ($P/A_0$) represents the engineering stress, with $A_0$ as the initial cross-sectional area of the tube. The horizontal axis ($\delta_\text{e}/L_\text{e}$) represents the overall axial Biot strain, with $L_\text{e}$ as the stereo-DIC gauge length and $\delta_\text{e}$ as the surface-averaged (axial) displacement of the tube measured by the stereo-DIC. Owing to careful gripping and alignments, consistent speckling, and precise stereo-DIC analyses, the two stress--strain curves from T1 and T2 measurements show satisfactory agreement. The superelastic stress--strain responses exhibit typical flat upper and lower plateaus, corresponding to the forward (A $\rightarrow$ M) and reverse (M $\rightarrow$ A) transformation stresses with a large stress hysteresis. Among all tested tubes, the 3.0/0.1/30 and 1.2/0.1/12 tubes exhibit the lowest stress-plateau of $\sim\!400$~MPa and the 2.5/0.28/8.93 and 2.5/0.13/19.23 tubes exhibit the highest stress-plateau of $\sim\!500$~MPa. With a stress hysteresis of $\sim\!243.5$~MPa, the 3.0/0.1/30 tube demonstrates the largest value, whereas the 1.2/0.2/6.0 tube demonstrates the smallest value, at $\sim\!187$~MPa.

The aforementioned differences in the superelastic stress--strain responses may originate from a combination of sources. The crystallographic texture induced during processing likely plays a role. Unfortunately, the absence of texture data at this stage precludes a definitive assessment. The DSC thermograms in Fig.~S1 reveal that the tubes possess noticeably different transformation temperatures. Such differences may result in distinct superelastic characteristics. It is well-known that increasing the test temperature $T_\text{test}$ leads to an upward shift of the stress plateaus~\citep{miyazaki1981transformation,shaw1995thermomechanical,ng2006stress,churchill2009tips,tsimpoukis2024thermomechanics}. Indeed, some of the observed variations in stress plateaus can be correlated with the temperature difference $T_\text{test}-T_\text{t}$, where $T_\text{t}$ is the equilibrium temperature and is here defined as the average of martensite start temperature $M_\text{s}$ and austenite finish temperature $A_\text{f}$ \citep{wayman1977equilibrium,salzbrenner1979thermodynamics}. For instance, the tubes 0.43/0.09/4.78, 0.8/0.13/6.15 and 1.2/0.1/12 exhibit a temperature difference $T_\text{test}-T_\text{t}$ of 75$^\circ$C, 64.6$^\circ$C and 53.1$^\circ$C, respectively, trending consistently with the observed decrease in the corresponding (upper and lower) stress plateaus. Also, the relatively low plateaus of the 3.0/0.1/30 tube can be attributed to its comparatively small $T_\text{test}-T_\text{t}=41.25^\circ$C. Note, however, that this criterion cannot explain all the stress discrepancies observed in Fig.~\ref{fig:mechsize}. Grain size is yet another factor to take into account. However, as shown in Fig.~S2 in Supplementary Data, variations in average grain sizes appear too subtle to exert a significant effect.

In what follows, we explore the evolution of transformation patterns and austenite--martensite front morphologies in the tubes and show their systematic dependence on geometrical factors $D$ and $D/t$. It is to be noted that certain features of the transformation pattern can be linked to the mechanical response, for instance, as shown later, tubes with lower transformation strains develop weaker localization effects. Nevertheless, there is no direct evidence connecting the specific front morphologies to the stress--strain characteristics. Based on this, we proceed with our investigation under the premise that the front morphology is mainly governed by $D$ and $D/t$.

\subsection{Size-dependent transformation patterns and front morphologies in tension}
\begin{figure}
    \centering
    \includegraphics[width=0.9\linewidth]{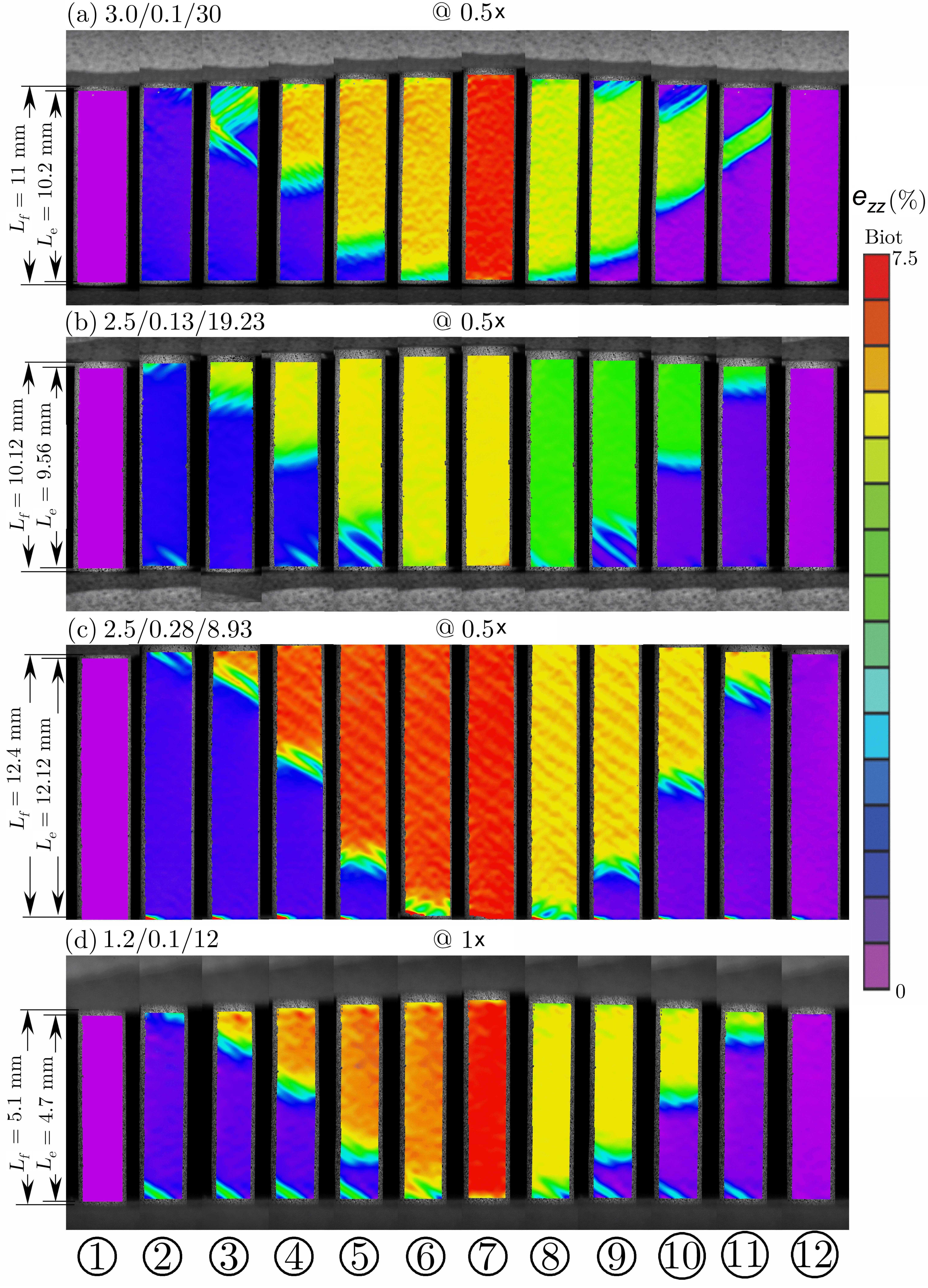}
    \caption{Surface evolution of axial strain field ($e_\text{zz}$) corresponding to the circled numbers \circled[10]{1} $\rightarrow$ \circled[10]{12} on tensile curves (Fig.~\ref{fig:mechsize}): (a) 3.0/0.1/30, T2, (b) 2.5/0.13/19.23, T1, (c) 2.5/0.28/8.93, T1, and (d) 1.2/0.1/12, T1.}
    \label{fig:DP1}
\end{figure}

\begin{figure}
    \centering
    \includegraphics[width=0.9\linewidth]{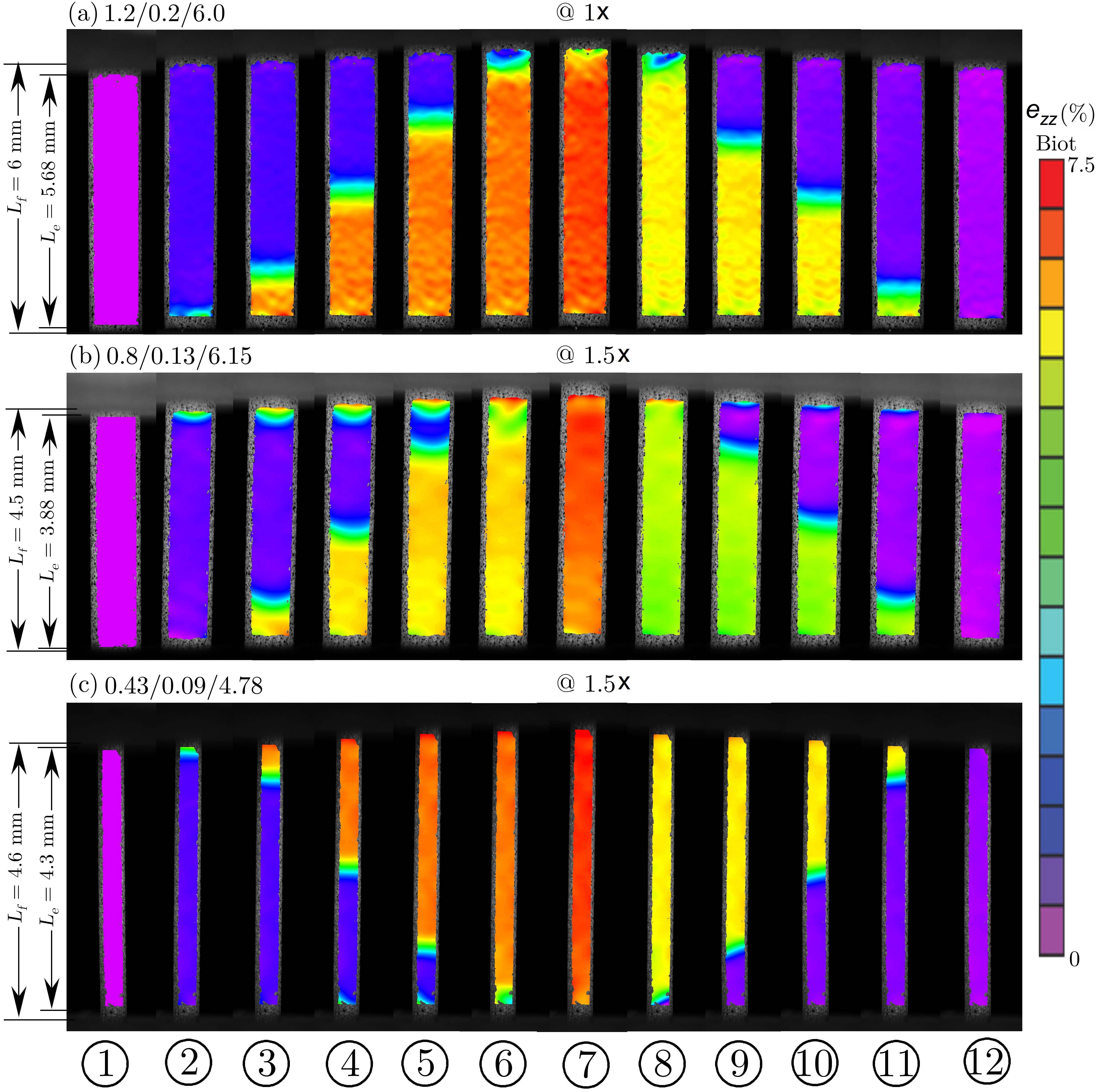}
    \caption{Surface evolution of axial strain field ($e_\text{zz}$) corresponding to the circled numbers \circled[10]{1} $\rightarrow$ \circled[10]{12} on tensile curves: (a) 1.2/0.2/6.0, (b) 0.8/0.13/6.15, and (c) 0.43/0.09/4.78 tubes, all pertaining to T1 tests.}
    \label{fig:DP2}
\end{figure}

\begin{figure}[!t]
    \centering
    \includegraphics[width=1\linewidth]{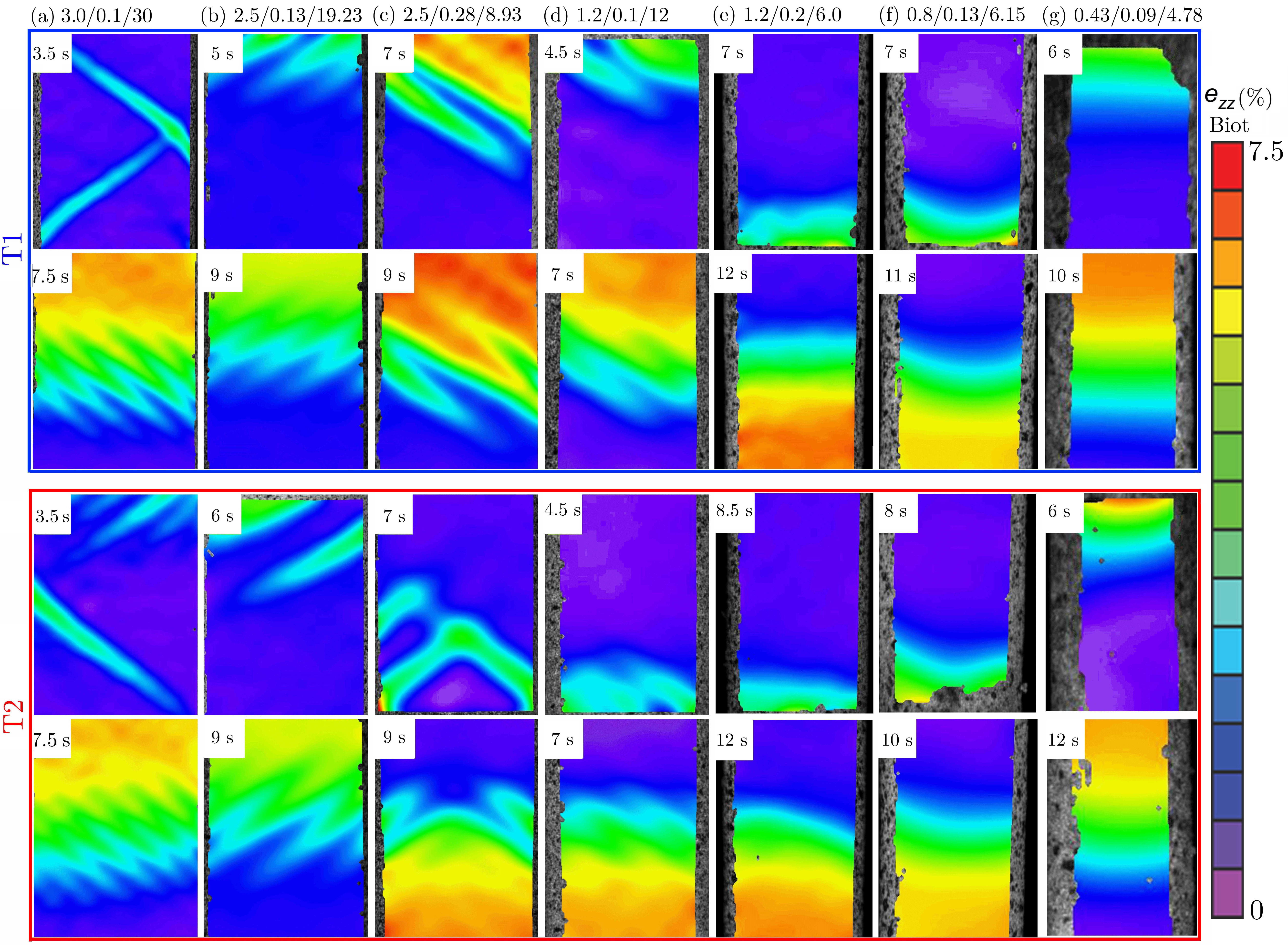}
    \caption{Magnified views of $e_\text{zz}$ at two selected time instants for T1 (blue box) and T2 (red box) tests. In each box, the upper set of images is representative of a time instant (strain) at the early stage of stress-induced PT (nucleation) and the lower set of images represents a time instant corresponding to the propagation of austenite--martensite interface.}
    \label{fig:DPZoom}
\end{figure}

The snapshots of the axial strain field ($e_\text{zz}$) corresponding to the circled numbers \circled[10]{1} $\rightarrow$ \circled[10]{12} on the stress--strain curves during loading (\circled[10]{1} $\rightarrow$ \circled[10]{7}) and unloading (\circled[10]{7} $\rightarrow$ \circled[10]{12}), measured on the front (viewable) side of the tubes are shown in Figs.~\ref{fig:DP1} and Fig.~\ref{fig:DP2}. The real-time evolution of the axial $e_\text{zz}$, hoop $e_{\theta\theta}$, and shear $e_{\text{z}\theta}$ strain fields, which are synchronized with the stress--strain responses are available as Supplementary Videos 1 to 14. To better illustrate the differences in transformation patterns and front morphologies among different tubes, enlarged views of the $e_\text{zz}$ field at two chosen time instants are presented in Fig.~\ref{fig:DPZoom}. The snapshots in blue and red boxes represent $e_\text{zz}$ field from T1 and T2 experiments, respectively. In each box, the upper-row images correspond to the time instants of early-stage transformation pattern evolution (i.e., the nucleation process), while the lower-row images correspond to the time instants when austenite--martensite front has formed and is propagating along the tube.

The figures and the Supplementary Videos clearly show that the stress-induced PT in the tubes with different $D$ and $D/t$ proceeds via nucleation and propagation of martensite domains exhibiting distinct patterning. In the T1 measurement on the 3.0/0.1/30 tube, the stress-induced PT starts with nucleation of two slim helical bands in a crossing manner (see Fig.~\ref{fig:DPZoom}(a) and Supplementary Video~1). In the T2 measurement, several bands nucleate in the upper grip, alongside a single band in the gauge section (see Fig.~\ref{fig:DPZoom}(a) and Supplementary Video~2). The PT then proceeds through concurrent formation of new helical bands at various sites within the gauge area, together with the spiral growth of the pre-existing bands. Ultimately, these bands merge and self-organize to create a rotating austenite--martensite front. At $\sim\!3.5\%$ strain (half of the length of the stress-plateau), the tube is divided into fully-transformed and un-transformed regions separated by a fingered (branched) front with strain variations across the interface. A comparison of the austenite--martensite front morphologies observed in the T1 and T2 measurements reveals that the fineness of the front is influenced by the number of the helical bands that nucleated at the onset of the stress-plateau, a factor that varies between the two measurements. In particular, the front morphology in T2 measurement is more refined than in T1; the fronts have about four and five fingers in T1 and T2 measurements, respectively. At strains larger than $\sim\! 4\%$ and towards the end of the stress-plateau (t~$\geq8$~s in Supplementary Video 2), the austenite--martensite front transitions into a more refined structure, driven by the emergence of fine crossing fingers. We posit that this front morphology refinement differs in nature from the morphology adaptation observed when two fronts come sufficiently close to each other to initiate merging, as can be clearly seen for instance in our 2.5/0.28/8.93 tube, see snapshots 5 and 6 in Fig.~\ref{fig:DP1}(c), and reported in the previous experiments on NiTi tubes \citep[e.g.,][]{bechle2016evolution}. In fact, this (early-appearing) morphology refinement may be associated with an effort to minimize the bending of the tube, akin to the criss-cross morphology observed in flat specimens \citep{jiang2017modeling,SHARIAT2022142774,rezaee2024modeling}.

Comparing the transformation patterns and front morphologies that evolve in the two $D = 2.5$~mm tubes with different $D/t$ ratios reveals a strong dependence of the related features to wall-thickness $t$ (see Fig.~\ref{fig:DP1}(b) and (c)). Note that due to a smaller transformation strain, the strain localization looks weaker in the thinner 2.5/0.13/19.23 tube (indicated by a yellowish $e_\text{zz}$ within the fully transformed region) as compared with the thicker 2.5/0.28/8.93 tube (indicated by a reddish $e_\text{zz}$ within the fully transformed region). Both T1 and T2 measurements on 2.5/0.13/19.23 tube revealed similar patterns and front morphologies (Fig.~\ref{fig:DPZoom}(b) and Supplementary Videos 3 and 4). The transformation patterns and front morphologies are qualitatively similar to those observed in 3.0/0.1/30 tube, nevertheless the helical martensite bands that nucleate in the $D=2.5$~mm tubes are fatter and the austenite--martensite front is coarser and have less fingers (four and three fingers in T1 and T2, respectively). On the other hand, two distinct morphologies are observed in T1 and T2 measurements in 2.5/0.28/8.93 tube. In the T1 test, (one or two) helical bands grow spirally into the austenite (see Supplementary Video 5), while in the T2 measurement a criss-cross pattern morphology appears, where two bands grow spirally into the austenite in opposite directions (see Supplementary Video 6). This pattern is similar to that reported by \citet{reedlunn2014tension} for a superelastic NiTi tube with comparable dimensions to this tube. It should be noted that the difference in the observed front morphologies between the T1 and T2 tests may also result from the fact that a limited portion of the tube surface is visible in the DIC setup. Thus, it is possible that the criss-cross pattern is also present in the T1 test, but remains undetected within the current field of view.

Analysis of the transformation patterns of the two $D = 2.5$~mm tubes shows that a smaller number of helical bands form in the thicker tube, and these bands are also fatter in comparison to those formed in the thinner tube. Consequently, the front that forms by coalescence of the bands is more refined in the thinner tube. Four/three fingers can be seen within the front in T1/T2 measurements in the thinner tube, while two/two fingers (spirally growing or criss-crossing) are observed in T1/T2 measurements in the thicker $D = 2.5$~mm tube.

The transformation patterns that evolve in the $D = 1.2$~mm tubes show marked differences. Both T1 and T2 measurements in the thinner 1.2/0.1/12 tube exhibit similar patterns to the previous tubes, i.e., one (or two) helical band(s) nucleate in the grip and grow spirally into the austenite (see Fig.~\ref{fig:DPZoom}(d) and Supplementary Videos 7 and 8, see also Fig.~\ref{fig:DP1}). Nevertheless, despite being thinner than the 2.5/0.28/8.93 tube, the front morphology is much less sharp and is considerably diffuse. This observation suggests that the characteristics of the transformation pattern and front morphologies in NiTi tubes is governed by both $D$ and $D/t$. In sharp contrast, the thicker 1.2/0.2/6.0 tube exhibits neither helical domain nucleation nor fingered front morphology, but rather a smooth but slightly curved transformation front. The two smaller 0.8/0.13/6.15 and 0.43/0.09/4.78 tubes also exhibit similar patterns and morphologies as the 1.2/0.2/6.0 tube, as can be seen in Fig.~\ref{fig:DP2} (see also Supplementary Videos 9 to 14). The observation of a cylindrical domain in the 0.43/0.09/4.78 tube is in-line with the optical observation by \citet{DONG20121063}.      

The real-time observations of the surface patterns and associated strain field evolutions presented above show that the front (gradually) transitions from a fingered morphology to a diffuse finger-less morphology as the tube becomes smaller and/or thicker. The key observations from the tensile experiments can be briefly summarized as follow:
\begin{enumerate}
    \item For a superelastic tube with a specific $D$, slim helical bands tend to nucleate at the onset of stress-plateau when the $D/t$ ratio is high. Moreover, the number of bands that nucleate increases with increasing $D/t$. 
    \item Spirally growing austenite--martensite fronts with fingered morphologies form in thin tubes. The front inherits its fineness from the number and morphology of the domains that nucleate at the onset of the stress-plateau. As such, there is a transition from finely-fingered austenite--martensite front to coarsely-fingered with decreasing $D/t$. 
    \item As $D/t$ decreases below some certain value, helical domain nucleation and fingered fronts are no longer observed. Flat and curved fronts form in such thick tubes.
    \item A diffuse spirally growing front is observed in a $D = 1.2$~mm tube with a $D/t = 12$~mm. Comparing the front morphology of this tube with larger but thicker tubes such as the 2.5/0.28/8.93 tube suggests that there is an interrelation between the transformation pattern and front features to both $D$ and $D/t$.   
\end{enumerate}

\subsection{Size-dependent moment--curvature bending responses}\label{sec:expbendres}
\begin{figure}[t!]
    \centering
    \includegraphics[width=0.5\linewidth]{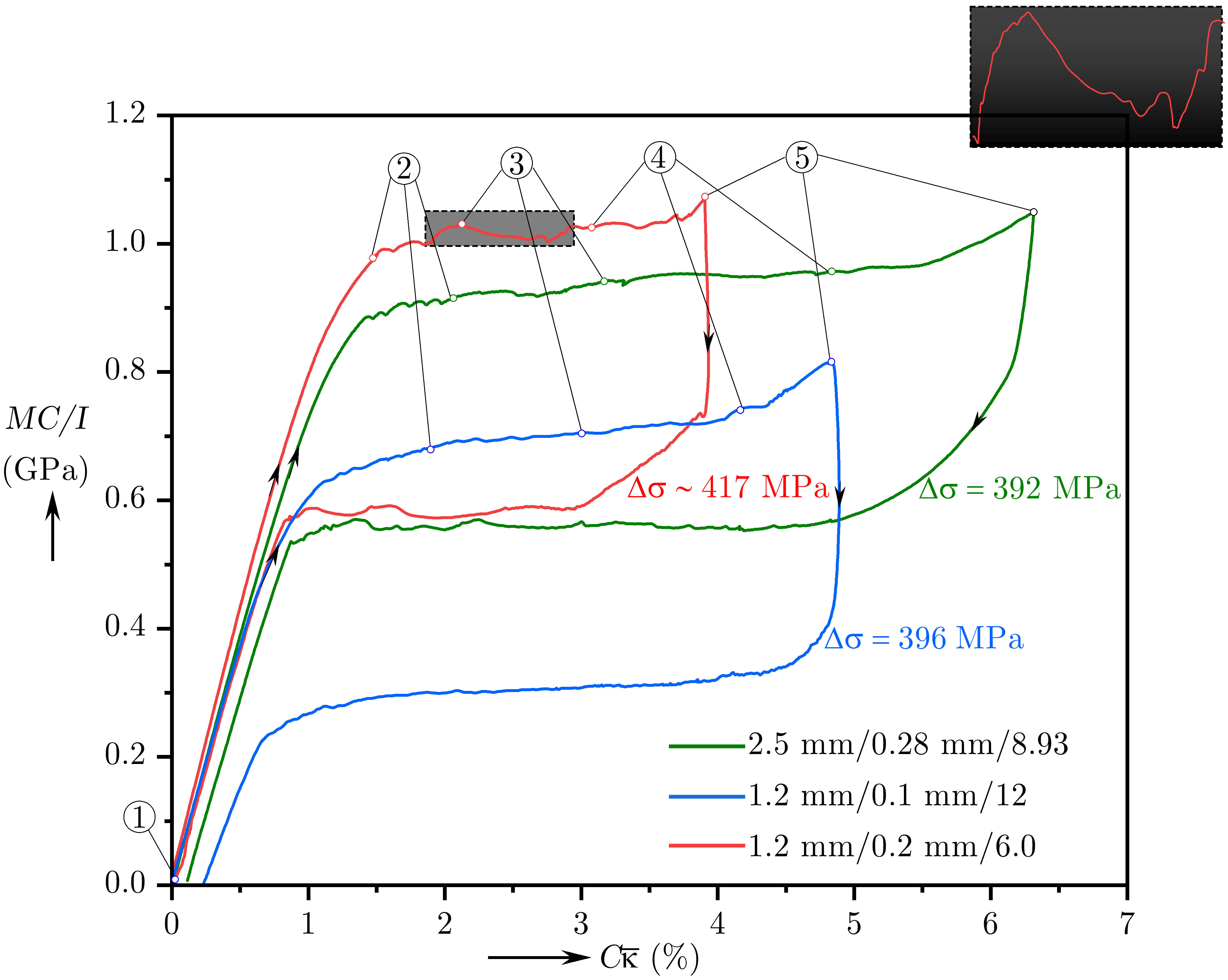}
    \caption{The bending moment--average curvature responses of 2.5/0.28/8.93, 1.2/0.1/12, and 1.2/0.2/6.0 tubes. The circled numbers correspond to snapshots of strain field images shown in Fig.~\ref{fig:DPBend}.}
    \label{fig:MomResp}
\end{figure}

Fig.~\ref{fig:MomResp} shows the global bending responses of the three tested tubes: 2.5/0.28/8.93, 1.2/0.1/12, and 1.2/0.2/6.0. The vertical axis ($MC/I$) represents the applied moment normalized by the second moment of inertia $I$ and $C=D/2$. The horizontal axis $C\,\overline{\kappa}$ represents the averaged curvature measured and calculated by stereo-DIC, as described in Section~\ref{sec:bend}. As shown by \citet{Berg1995}, these classical formulas ($MC/I$ and $C\overline{\kappa}$) can satisfactorily represent the bending responses of round NiTi specimens with different sizes without the intrusion of geometrical factors. In analogy to uniaxial stress--strain curves, the tubes sustain an initial elastic loading prior to stress-induced PT. The average curvature (strain) corresponding to the initiation of PT is $\sim\! 1.4$~\%, 1\%, and 1.7\% in the 2.5/0.28/8.93, 1.2/0.1/12, and 1.2/0.2/6.0 tubes, respectively. These values are larger than the measured elastic strain limits of the tubes under tension (see Fig.~\ref{fig:mechsize}), and are in line with the measurements of~\citet{reedlunn2014tension}, in which a slightly larger strain limit of $\sim\! 2\%$ in a 3.18/0.318/10 tube was identified. The slight difference can be attributed to the difference in the design and dimensions of the bending fixtures. However, by comparing the elastic limits of the 1.2/0.1/12 and 1.2/0.2/6.0 tubes (1\% vs. 1.7\%) and considering the same tensile elastic limit of both tubes ($\sim\! 0.8$~\%), it is noted that with decreasing $D/t$ the elastic strain limit increases, i.e., thicker tubes can sustain larger elastic strains under bending deformation.

Above the elastic limit, the stress--induced PT kicks in and the moment--curvature responses exhibit plateau-like superelasticity. In the transformation regime, discrepancies between the bending moment--curvature responses are found. We first note that, as was shown in Fig.~\ref{fig:mechsize}, the lengths of the tensile stress-plateaus of the three tested tubes are almost identical ($\sim\!6.5$~\%). Thus, one expects similar lengths of the moment-plateaus under bending. However, it is clearly seen that the tubes exhibit different lengths of the moment-plateaus, with the largest (2.5/0.28/8.93) tube having the longest plateau. Second, one expects the 2.5/0.28/8.93 tube with the highest tensile plateau-stress of $\sim\!500$~MPa to exhibit the highest moment-plateau under bending as well. However, the 1.2/0.2/6.0 tube exhibits the highest moment-plateau of $\sim\!1$~GPa. At the same time, given the $\sim\!80$~MPa difference in the tensile plateau-stresses of the 1.2/0.1/12 and 1.2/0.2/6.0 tubes, the $\sim\!300$~MPa difference in the corresponding moment-plateaus also suggests that the global bending responses seem to be strongly $D/t$-dependent. Similar discrepancies can be observed in the stress hysteresis of the tubes. 

Moreover, in the plateau regime, the overall moment--curvature responses have a slightly positive slope for all the tubes i.e., the plateau is not perfectly horizontal. Also, small moment serrations are noticed. The serrations are stronger in the 1.2/0.2/6.0 tube than the other two tubes and the plateau is smoother in the thinnest (1.2/0.1/12) tube. The interesting observation here is that in the thickest (1.2/0.2/6.0) tube, the moment actually decreases in specific strain segments in the plateau (see the inset in the upper-right corner of Fig.~\ref{fig:MomResp}). This behavior is reminiscent of the bending response of the steels that exhibit curvature propagation \citep{KYRIAKIDES20083074}, although the effect here is not as obvious. In the initial stage of the unloading process, the load decreases steeply in the 2.5/0.28/8.93 tube. In the 1.2/0.1/12 and 1.2/0.2/6.0 tubes, however, the initial unloading segment is visibly vertical. With further unloading, the moment decreases monotonically until reverse moment-plateau is reached and eventually the curvature is mostly recovered. The fact that the global moment--curvature response of the 2.5/0.28/8.93 tube resembles that reported by \citet{reedlunn2014tension} and the response of the 1.2/0.2/6.0 tube resembles that reported by \citet{bechle2014localization} confirms that unlike the global tensile response, the bending response is indeed affected by tube's dimension. These differences originate from the role that the dimension and associated geometrical constraints play on nucleation, propagation, and merging of martensite domains with different morphologies.

\subsection{Size-dependent transformation patterns under pure bending}\label{sec:expbendresTP}
\begin{figure}
    \centering
    \includegraphics[width=1\linewidth]{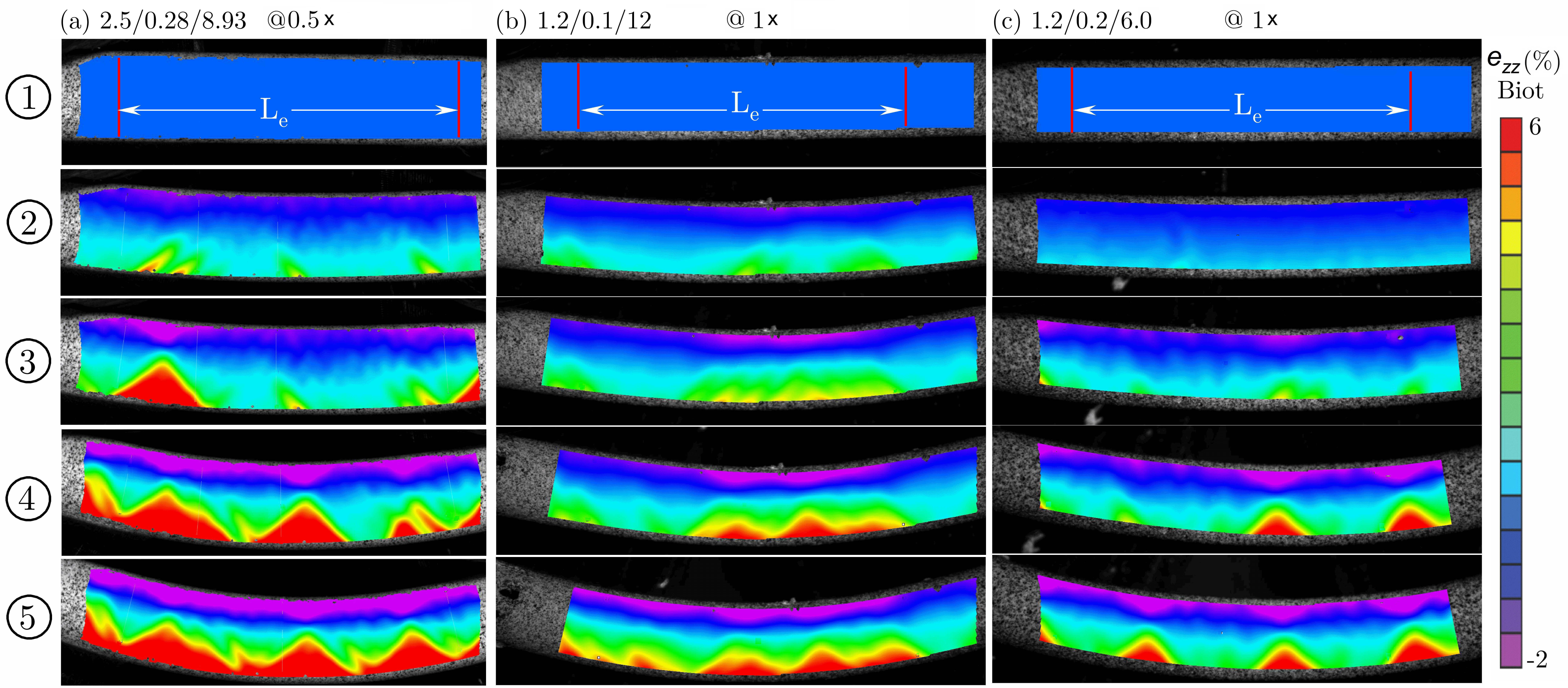}
    \caption{Snapshots of the axial strain fields ($e_\text{zz}$) corresponding to the circled numbers \circled[10]{1} $\rightarrow$ \circled[10]{5} on moment--curvature curves shown in Fig.~\ref{fig:MomResp}.}
    \label{fig:DPBend}
\end{figure}

The snapshots of the axial strain fields ($e_\text{zz}$) corresponding to the circled numbers \circled[10]{1} $\rightarrow$ \circled[10]{5} on moment--curvature curves (Fig.~\ref{fig:MomResp}) are shown in Fig.~\ref{fig:DPBend}. Unloading snapshots are not shown here, however, the full loading--unloading evolution of the strain fields can be seen in Supplementary Videos 15 to 17. The two thick red lines in snapshots \circled[10]{1} are the positions where the end rotations ($\varphi_1$ and $\varphi_2$) are measured, which are used to calculate the average curvature $\overline{\kappa}$. The distance between the two lines defines $L_\text{e}$. 

In the initial (nearly) linear stage of the moment--curvature response, the tensile and compressive strain fields develop in similar fashion for all the tubes. Once the strain exceeds the elastic limit and reaches moment-plateau, the tensile side of the tubes start transforming to martensite with different pattern morphologies. In the 2.5/0.28/8.93 tube, two sharp fingered martensite bands nucleate on the left-hand side, one in the middle, and another one on the right-hand side, as shown in \circled[10]{2} in Fig.~\ref{fig:DPBend}(a). At \circled[10]{3}, deformation is localized on the left-hand side where the two fingers grow (mainly laterally) and merge to form the first wedge. This process is accompanied by the transformation of the tube at its far right-end and a temporary stall of the fingers formed in the middle and right-hand side of the tube. With further straining, the stalled fingers resume growing and form wedges. Hence, at the end of the moment-plateau, corresponding to \circled[10]{5} in Fig.~\ref{fig:DPBend}(a), three wedges are formed. The moment--curvature response as well as the transformation pattern that evolves in this tube are nearly identical to those observed in \citet{reedlunn2013superelastic, watkins2018uniaxial}, however, similar to the behavior under tension, the fingered bands that form here are less sharp. 

The strain field evolution of the 1.2/0.1/12 and 1.2/0.2/6.0 tubes are shown in Fig.~\ref{fig:DPBend}(b) and (c), see also Supplementary Videos 16 and 17. The first difference is that the fingers that form in the 2.5/0.28/8.93 tube are slimmer, sharper, and they are clearly distinguishable. In contrast, the fingers formed in the two smaller tubes are fatter, much more diffuse and not clearly distinguishable. A close inspection of the transformation pattern shows that more fingered bands nucleate in the 1.2/0.1/12 tube; almost nine burst-like events are noted in the course of loading. With further bending (\circled[10]{4}), the two tubes develop two distinct patterns. More dispersed transformation regions appear in the 1.2/0.1/12 tube, which create a nugget-shaped high-strain morphology in the middle of the tube. Although not presented, this pattern was repeatedly observed across multiple measurements even with different $W$ distances. In stark contrast, wedges continuously evolve in the regions where very diffuse fingers had previously appeared in the 1.2/0.2/6.0 tube. With further straining in this tube, hump-like martensite regions evolve at the left-hand and right-hand side of the wedges (notice the greenish area surrounding the wedges), making a crown-shaped high-strain morphology. The crown-shaped morphology in the 1.2/0.2/6.0 tube is also captured in the modeling, as will be discussed later in Section~\ref{sec:analbend}. 

Let us now elaborate further on the crown-shaped morphology. As observed in \citet{reedlunn2014tension}, towards the end of the moment-plateau, when the wedges fully develop in the tensile region, `baby' fingers nucleate in the inter-wedge regions to accommodate further straining in the tube (see snapshots \circled[10]{5}--\circled[10]{8} in Fig.~11(a) of \citet{reedlunn2014tension}). This behavior is also observed in our 2.5/0.28/8.93 tube (see snapshots \circled[10]{4} and \circled[10]{5} in Fig.~\ref{fig:DPBend}(a) or time duration 11--13~s in Supplementary Video 15). The formation of the crown-shaped morphology in the 1.2/0.2/6.0 tube is actually a similar response to accommodate the applied bending deformation. However, due to the geometrical constraints and presence of strain gradients in a small region, the wedges cannot fully develop in the small and thick 1.2/0.2/6.0 tube. As such, bending deformation is accommodated by a more uniform evolution of the strain in the inter-wedge regions, which results in the crown-shaped morphology. Note that even in the thinner 1.2/0.1/12 tube, we always observed regions with much less developed strain-fields, which confirms that the progression of stress-induced PT in smaller diameter tubes is intrinsically more difficult due to geometrical constraints.

The real-time observations of the strain field evolutions during pure bending experiments on NiTi tubes with different sizes demonstrate a strong dependence of the transformation pattern to $D$ and $D/t$. This size-dependent behavior, captured for the first time in this work, is also reflected in the corresponding moment--curvature responses. The key observations of the bending experiments can be briefly summarized as follow:
\begin{enumerate}
    \item Unlike uniaxial tension, the main characteristics of moment--average curvature superelasticity, such as elastic strain limit, moment-plateau level, length of the moment-plateau, smoothness, and stress hysteresis are noticeably affected by tube dimensions. 
    \item A transition from sharp fingered martensite bands (in large tubes) to fat and diffuse bands (in small tubes) is observed. 
    \item Sharp fingers grow and form wedges, which develop with further loading and accommodate the applied large deformations in large tubes. In small and thick tubes, diffuse fingers continuously evolve into wedges. The wedges, however, cannot fully develop, leaving quasi-dead-zone (i.e., minimally transformed) inter-wedge regions. With further bending, the strain uniformly and continuously evolves in these inter-wedge regions. 
 
\end{enumerate}

\section{Modeling}\label{sec:anal}
In this section, the experimental findings on the dimension-dependent transformation patterns of NiTi tubes are validated and examined via finite-element analysis. Our main aim is to elucidate the morphological transitions across different tube dimensions by investigating the interplay between the bulk energy of the tube and the interfacial (gradient) energy of the macroscopic transformation front (austenite--martensite interfaces). This interplay arises from the energy minimization mechanism and dictates the equilibrium morphology of the evolved martensite domain. We employ a state-of-the-art gradient-enhanced model of superelasticity for our investigation. The details of the model formulation and its subsequent refinements can be found in \citet{RezaeeStupkiewicz2018Gradient,Rezaee2020Gradient,rezaee2024modeling}.

\subsection{The gradient-enhanced model of superelasticity}\label{sec:model}
The model is formulated within the finite-deformation theory and adheres to the incremental energy minimization approach. Its constitutive description is thus defined by two (local) potentials: the Helmholtz free energy density $\phi$ and the dissipation potential $\Delta D$. A global incremental potential $\Pi$ is constructed from the global counterparts of the two potentials and the solution of the problem is sought by minimizing $\Pi$ with respect to the problem unknowns, 
\begin{equation}\label{eq:min}
    \Pi= \Delta \Phi + \Delta \mathcal{D} \rightarrow \; \min_{\substack{\bm{\varphi},\bar{\bm{e}}^\text{t},\eta}}
\end{equation}
where $\Phi=\int_\Omega \phi\,\rd V$ and $\Delta \mathcal{D}=\int_\Omega \Delta D\,\rd V$ denote the global free energy functional and the global incremental dissipation potential, respectively. To fix the attention, the potential of external loads has not been included in the minimization problem~\eqref{eq:min}.

As appears in Eq.~\eqref{eq:min}, the minimization is performed with respect to three unknowns, namely $\bm{\varphi}$, $\bar{\bm{e}}^\text{t}$ and $\eta$, which are briefly introduced here. Central to the finite-deformation theory is the deformation gradient $\bm{F}$, defined as $\bm{F}=\nabla \bm{\varphi}$, where $\bm{\varphi}$ represents the mapping that links the material point from the reference configuration to the current (deformed) configuration, and the symbol $\nabla$ denotes the spatial gradient with respect to the reference configuration. The mapping  $\bm{\varphi}$ constitutes the first unknown in the minimization problem~\eqref{eq:min}. The deformation gradient $\bm{F}$ is multiplicatively decomposed into an elastic component $\bm{F}^\text{e}$ and a transformation component $\bm{F}^\text{t}$. Upon applying the polar decomposition, the transformation part $\bm{F}^\text{t}$ is further decomposed into a stretch tensor $\bm{U}^\text{t}$ and a rotation tensor $\bm{R}^\text{t}$. The stretch tensor $\bm{U}^\text{t}$ is then defined as an exponential function of the (logarithmic) transformation strain tensor $\bm{e}^\text{t}$ \citep{StupkiewiczPetryk2013Robust}, while the rotation $\bm{R}^\text{t}$ can remain unspecified in the case of isotropic elasticity, as assumed here. These kinematic relations are given below,
\begin{equation}
    \bm{F}=\bm{F}^\text{e}\bm{F}^\text{t}, \quad \bm{F}^\text{t}=\bm{R}^\text{t}\bm{U}^\text{t}, \quad \bm{U}^\text{t}= \exp \bm{e}^\text{t}.
\end{equation}
The transformation strain tensor $\bm{e}^\text{t}$ is assumed as the product of two internal variables: the limit transformation strain $\bar{\bm{e}}^\text{t}$ and the volume fraction of martensite $\eta$, i.e., $\bm{e}^\text{t}=\eta \bar{\bm{e}}^\text{t}$. These two variables collectively characterize the transformation state of the material and form the remaining unknowns in the minimization problem~\eqref{eq:min}. Specifically, the volume fraction $\eta$ quantifies the proportion of martensite within the material and is constrained by $0 \leq \eta \leq 1$, where $\eta=0$ and $\eta=1$ correspond to pure austenite and pure martensite phases, respectively. Meanwhile, the limit transformation strain $\bar{\bm{e}}^\text{t}$, which determines the direction of the transformation path, is bounded by the surface $g(\bar{\bm{e}}^\text{t})=0$. There exists some latitude in defining the surface $g(\bar{\bm{e}}^\text{t})=0$, which is leveraged to incorporate key features of polycrystalline (textured) NiTi, namely tension--compression asymmetry and transverse isotropy. The explicit form of the surface $g(\bar{\bm{e}}^\text{t})=0$ is not provided here; details can be found in \citet{StupkiewiczPetryk2013Robust}. 

We now return to the definition of the local potentials $\phi$ and $\Delta D$. The Helmholtz free energy function $\phi$ consists of two principal contributions: the bulk energy and the gradient energy,
\begin{equation}\label{eq:en:bg}
    \phi=\phi_\text{bulk}+\phi_\text{grad}.
\end{equation}
Here, to maintain the generality of the framework, the variables are not indicated as the explicit arguments of the energy functions. The bulk energy $\phi_\text{bulk}$ encompasses the chemical energy of phase transformation, $\phi_\text{chem}$, the elastic strain energy, $\phi_\text{el}$, and the so-called austenite--martensite interaction energy, $\phi_\text{int}$, so that
\begin{equation}\label{eq:en:b}
\phi_\text{bulk}=\phi_\text{chem}+\phi_\text{el}+\phi_\text{int}.
\end{equation}
The bulk energy components are formulated in standard forms. The chemical energy $\phi_\text{chem}$ is defined as the weighted average of the (stress-free) free energy densities of pure austenite and martensite phases,
\begin{equation}\label{eq:chem}
        \phi_\text{chem}(\eta)=(1-\eta) \phi_0^\text{a} + \eta \phi_0^\text{m}=\phi_0^\text{a}+ \Delta \phi_0 \eta,
    \end{equation}
    with $\Delta \phi_0$ denoting the chemical energy of transformation. An isotropic elastic strain energy $\phi_\text{el}$ is adopted in the following neo-Hookean form,
    \begin{equation}\label{eq:neohook}
        \phi_\text{el}(\bm{F},\bar{\bm{e}}^\text{t},\eta)=\frac{1}{2} \mu(\eta) (\tr(\hat{\bm{b}}^\text{e})-3)+\frac{1}{4} \kappa (\det (\bm{b}^\text{e})-1-\log(\det(\bm{b}^\text{e}))),
    \end{equation}
    with $\bm{b}^\text{e}=\bm{F}^\text{e}(\bm{F}^\text{e})^\text{T}$ as the left Cauchy--Green tensor and $\hat{\bm{b}}^\text{e}=(\det(\bm{b}^\text{e}))^{-1/3}\,\bm{b}^\text{e}$ as its volume-preserving part. While the bulk modulus $\kappa$ is considered to be independent of the volume fraction $\eta$, the shear modulus $\mu$ is $\eta$-dependent and its value is obtained by a Reuss averaging of the shear moduli of pure austenite ($\mu_\text{a}$) and martensite ($\mu_\text{m}$) phases, $1/\mu(\eta)=(1-\eta)/\mu_\text{a}+\eta/\mu_\text{m}$. The interaction energy $\phi_\text{int}$ is adopted as a quadratic function of the volume fraction $\eta$, 
     \begin{equation}\label{eq:phiint}
        \phi_\text{int}(\bar{\bm{e}}^\text{t},\eta)=\frac{1}{2}H(\bar{\bm{e}}^\text{t}) \eta^2.
    \end{equation}
    Note that $\phi_\text{int}$ characterizes the mechanical response within the transformation regime, featuring a softening-type or a hardening-type behavior. In fact, the parameter $H$ in Eq.~\eqref{eq:phiint} is formulated as a linear function of the norm of the transformation strain $\bar{\bm{e}}^\text{t}$, thereby making the response loading-mode dependent. Consequently, a softening-type response in tension and a hardening-type response in compression can be obtained. This is an inherent feature of the material that plays a crucial role for the simulation of NiTi tubes subjected to bending wherein both tensile and compressive stresses are simultaneously in action \citep{jiang2017bending,rezaee2024modeling}.

Having the bulk energy components specified, along with an appropriate dissipation potential (to be introduced later), the intrinsic material response can be fully described. The present formulation yields a trilinear (elastic--transformation--elastic) material response with abrupt transitions between the elastic and transformation branches. In the case of a softening-type material response (where $H<0$), the boundary-value problem becomes ill-posed, rendering the solution unreliable and severely mesh-dependent, unless the problem is adequately regularized. To address this issue, as is customary, the free energy function is enhanced by gradient terms; suitable to the present framework, by the gradient of the volume fraction $\eta$. Accordingly, the gradient energy $\phi_\text{grad}$ is introduced in the following form,
    \begin{equation}\label{eq:graden}
    \phi_\text{grad}(\nabla \eta)=\frac{1}{2} G \nabla \eta \cdot \nabla \eta,
\end{equation}
with $G>0$ as the gradient energy parameter. The inclusion of the gradient energy $\phi_\text{grad}$ brings on a characteristic length, $\lambda=\pi \sqrt{-G/H}$ (for $H<0$), to the otherwise ill-posed problem, thereby gives rise to a size-dependent transformation behavior. The characteristic length $\lambda$ represents the theoretical thickness of the austenite--martensite diffuse interface and can be used as a reference to select a proper value for the gradient energy parameter $G$ \citep{RezaeeStupkiewicz2018Gradient}, see more details in Section.~S2.2 in Supplementary Data.

Finally, the model postulates a rate-independent dissipation potential, expressed in the incremental form as
\begin{equation}\label{eq:diss}
    \Delta D(\Delta \bm{e}^\text{t},\Delta \eta)=f_\text{c} |\Delta \eta|+ \frac{f_\text{r}}{\Vert \bar{\bm{e}}^\text{t}\Vert} \Vert \Delta \bm{e}^\text{t} \Vert, \quad \Vert \Delta \bm{e}^\text{t} \Vert=\sqrt{\Delta \bm{e}^\text{t} \cdot \Delta \bm{e}^\text{t}},
\end{equation}
where $\Delta \bm{e}^\text{t}=\bm{e}^\text{t}-\bm{e}^\text{t}_n$ and $\Delta \eta=\eta-\eta_n$, with the subscript $n$ denoting quantities from the previous time step. The parameters $f_\text{c}$ and $f_\text{r}$ are the critical driving forces associated with the phase transformation and reorientation mechanisms, respectively, and jointly define the width of the hysteresis loop. The finite-element implementation of the model is provided in the Supplementary Data, see Section~S2.1.

\subsection{NiTi tubes under quasi-static tension}\label{sec:analten}
The present study aims to investigate the variations in the free energy contributions of NiTi tubes and the resulting morphological transitions in response to changes in tube size and thickness. To this end, we first analyze geometrically self-similar tubes, with fixed diameter-to-thickness and diameter-to-length ratios of $D/t=25$ and $D/L=0.4$, by varying the tube diameter over the range $0.43~\text{mm} \leq D \leq 3~\text{mm}$. This enables us to focus solely on size effects and eliminate the intervention of other geometrical parameters. Next, we turn our attention to the role of tube thickness in driving the morphological transitions. We thus fix the tube diameter and length at $D=2.5$~mm and $L=6.25$~mm and vary the tube thickness over the range $0.025~\text{mm} \leq t \leq 0.28~\text{mm}$, corresponding to the diameter-to-thickness ratios of $8.9 \leq D/t \leq 100$. The selected ranges are sufficiently broad to encompass the geometries tested in the experiment, see Section~\ref{sec:tubes}.

To ensure that the results remain free from artifacts, a mesh sensitivity analysis is conducted for each geometry under consideration. The reported results thus correspond to the finest mesh density, beyond which no further improvements are detectable. Note also that, to eliminate the mesh bias and adequately capture the out-of-plane bending, nearly equiaxed elements are employed. This results in pretty high computational costs, especially for tubes with a low diameter-to-thickness ratio. Therefore, to mitigate this, the diameter-to-length ratio of the tubes is limited to $D/L=0.4$. At the same time, only half of the tube's cross section is simulated, with proper symmetry conditions imposed along the symmetry plane. We use simplified boundary conditions, fixing the axial displacement of the bottom-section nodes to zero, while prescribing the axial displacement of the top-section nodes. Moreover, the rigid-body movement of the tube is prevented by appropriately constraining the lateral displacements. 

With the aim to isolate the influence of geometrical factors and more effectively elucidate their role in shaping the energy balance of the tubes, the numerical investigation is conducted under the assumption that both the material response and the gradient energy parameter $G$ are identical across all tube geometries. Consequently, an identical set of material parameters is employed throughout all simulations, see Section~S2.2 in Supplementary Data. The analysis is focused on evaluating the bulk and gradient energy contributions at an overall (engineering) axial strain of $4\%$, corresponding to the instant where the transformation front is positioned near the mid-length of the tube. To enable direct comparisons, it is imperative that the transformation front propagates along the same direction in all the tubes, thereby ending up consistently at the same location. This is achieved by introducing a geometric imperfection, as a surface indent, near the bottom section to trigger the localization. The energy of the tube is then assessed using the following energy densities,
\begin{equation}\label{eq:enint}
    \langle \Phi_k \rangle=\left.\!\langle \Phi_k \rangle \right|_z =\frac{1}{A_\text{c}} \int_{\Gamma_\text{c}} \phi_k\, \rd A_\text{c}, \quad \{ \Phi_k \} =\frac{1}{V} \int_\Omega \phi_k\, \rd V, \quad k \in \{ \text{bulk}, \text{grad}\},
\end{equation}
where $A_\text{c}$ is the cross-sectional area and $V$ is the total volume. The quantities $\langle \Phi_k \rangle$ and $ \{ \Phi_k \}$ represent the average energy density over the cross-sectional area and the overall average energy density, respectively. Notably, $\langle \Phi_k \rangle$ can be evaluated at each axial position $z$, and provides the distribution of energy density along the tube axis.

For each tube under investigation, a reference energy state is established, corresponding to an axisymmetric tube configuration in which the formation of a patterned (fingered) front is precluded, and, regardless of the size and thickness, the front consistently adopts a flat ring-like (finger-less) morphology. The axisymmetric configuration serves as a baseline to highlight how the bulk and gradient energy contributions vary and interact with each other as a result of the formation of the fingered morphology.

\begin{figure}
    \hspace*{0pt}
    \includegraphics[width=1\textwidth]{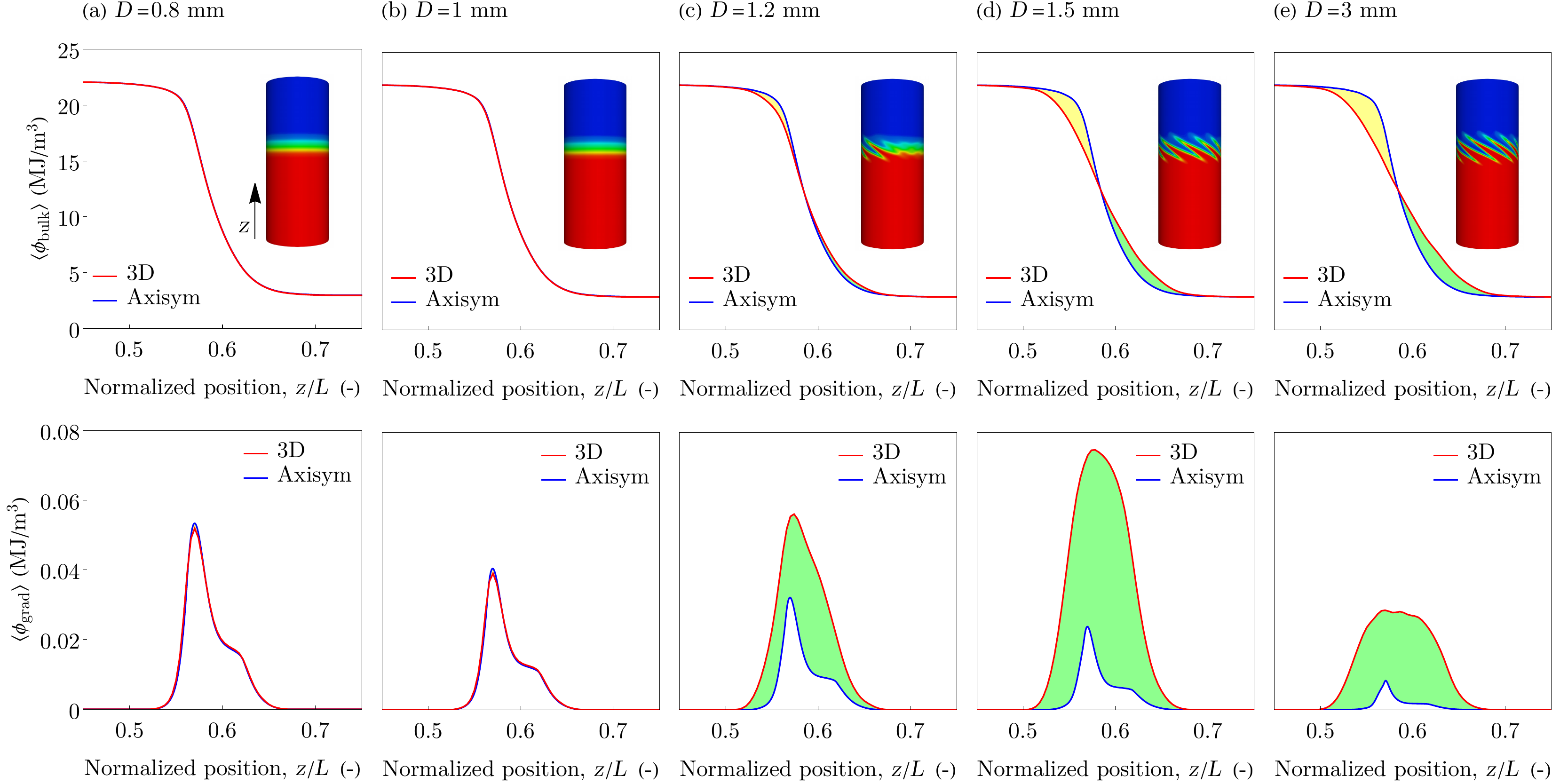}
    \caption{Profiles of the average bulk and gradient energy densities for varying tube diameter $D$ (recall that the ratios $D/t=25$ and $D/L=0.4$ are fixed). In the graphs, greenish shading indicates the regions where the energy density of the full tube exceeds that of the axisymmetric reference, while yellowish shading indicates the opposite.}
    \label{fig:SimSizeSum}
\end{figure}

Fig.~\ref{fig:SimSizeSum} presents the profiles of the average bulk and gradient energy densities for various tube diameters $D$. The transformation patterns are included as insets within the bulk energy panels and are represented by the distribution of the martensite volume fraction $\eta$. Several key observations emerge from Fig.~\ref{fig:SimSizeSum}:
\begin{enumerate}
    \item For larger tubes ($D=1.5$~mm and $D=3$~mm), the transformation front exhibits a fingered morphology, characterized by 5--6 distinct fingers in the simulated half-tube. On the contrary, smaller tubes ($D=0.8$~mm and $D=1$~mm) develop a ring-like front devoid of fingers. The tube with $D=1.2$~mm represents a transitional state, featuring a front that appears as a mixture of fingered and ring-like morphologies, with 2 clearly distinguishable fingers. The simulated front morphologies demonstrate a qualitatively close resemblance to the morphologies captured in the experiment (see Fig.~\ref{fig:DPZoom}).
    \item Obviously, the change in the front morphology is associated with a change in the energy state of the system. This is manifested by the growing deviation between the bulk energy density, $\langle \Phi_\text{bulk} \rangle$, of the full tube and its axisymmetric reference for larger $D$ (the deviations are highlighted by yellowish and greenish areas, as explained in the caption of Fig.~\ref{fig:SimSizeSum}). Such a growing deviation is expected, as the front morphology of the full tube diverges more markedly from the reference ring-like configuration as $D$ becomes larger.
    \item The gradient energy density, $\langle \Phi_\text{grad} \rangle$, shows a more dynamic behavior. When the front morphology remains unchanged with increasing $D$, $\langle \Phi_\text{grad} \rangle$ diminishes, indicating less pronounced gradient energy effects for larger volume. This trend is consistently observed in the axisymmetric tube, as well as in the full tube in two cases: between $D=0.8$~mm and $D=1$~mm, and between $D=1.5$~mm and $D=3$~mm. However, when the front morphology changes with increasing $D$, the gradient energy experiences an escalation, see the changes from $D=1$~mm to $D=1.5$~mm. This escalation stems from the increase in the interfacial area, as a consequence of the formation of fingers. 
\end{enumerate}

\begin{figure}
    \hspace{0pt}
    \includegraphics[width=1\textwidth]{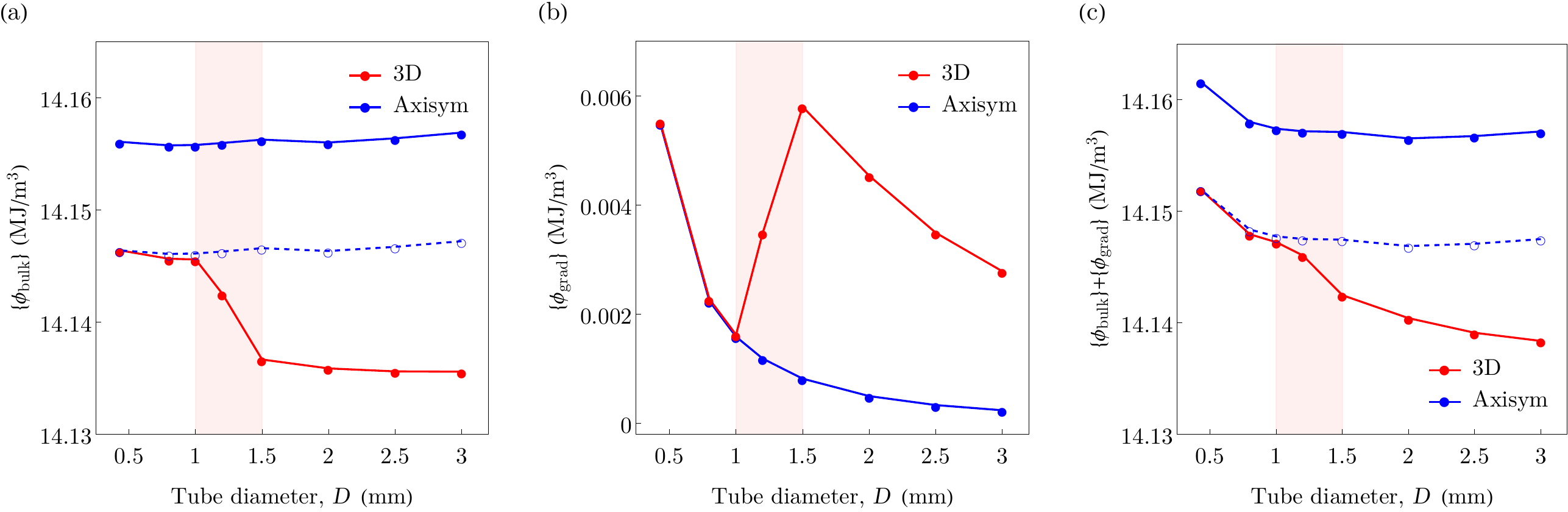}
    \caption{The overall average energy densities as a function of the tube diameter $D$: (a) the bulk energy density $\{ \phi_\text{bulk} \}$, (b) the gradient energy density $\{ \phi_\text{grad} \}$, and (c) the sum of the two contributions. The dashed blue lines in panels (a) and (c) represent the adjusted energy densities of the axisymmetric tube, see the text for further details.}
    \label{fig:SimSizeEn}
\end{figure}

Fig.~\ref{fig:SimSizeEn} depicts the overall average energy densities, as defined in Eq.~\eqref{eq:enint}$_2$, as a function of the tube diameter $D$. It provides a comprehensive view on size effects by highlighting more clearly the interplay between the bulk and gradient energy contributions and their correlation with the observed morphological changes. As already observed in Fig.~\ref{fig:SimSizeSum}, Fig.~\ref{fig:SimSizeEn}(b) shows, from a global perspective: (1) the close alignment between the gradient energies of the full and axisymmetric tubes for small $D$, (2) the steady asymptotic decline of the gradient energy in the axisymmetric tube as $D$ increases, (3) the sharp gradient energy escalation of the full tube within the transitional region (see the shaded area in Fig.~\ref{fig:SimSizeEn}), and (4) the continuation of its declining trend beyond the transitional region. Meanwhile, the bulk energy density, $\{ \Phi_\text{bulk} \}$, follows a relatively mild declining trend on either side of the transitional region, and is marked by a sharp drop within the transitional region. Summing up the two contributions gives rise to an overall declining trend, with a noticeably steeper slope within the transitional region.

The size-dependent nature of the gradient energy is, in fact, the main driver of the size effects observed in this study. As the tube size decreases, the gradient energy becomes increasingly dominant, effectively governing the energy minimization process. Therefore, a front with the smallest interfacial area, i.e., the ring-like front, is energetically preferred. Conversely, as the tube size increases, the gradient energy effects diminish, shifting the energy minimization process towards favoring front morphologies that reduce the bulk energy of the system, i.e., the fingered front, even at the expense of increasing the gradient energy. Recall that the interfacial energy of the front is governed by the scale-dependent parameter $G$, see Eq.~\eqref{eq:graden}. The parameter $G$ plays a critical role in modulating the relative contribution of the gradient energy compared to the bulk energy. Notably, the variations in the bulk and gradient energy densities observed in Fig.~\ref{fig:SimSizeSum} and Fig.~\ref{fig:SimSizeEn} can be reproduced by adjusting the value of $G$ while keeping the tube size fixed \citep{he2009effects}. Specifically, increasing/decreasing $G$ is equivalent to decreasing/increasing tube size.

As a side note, a minor discrepancy can be observed between the overall bulk energy densities, and consequently between the total energy densities, of the full and axisymmetric tubes for the small tube diameters, i.e., for $D \leq 1$~mm, despite the fact that the corresponding energy profiles agree perfectly well within the region of interest, as shown in Fig.~\ref{fig:SimSizeSum}. We attribute this discrepancy to two factors. First, we employ different finite-element technologies for the two problems at hand: TSCG12 element for the displacement field in the 3D setup \citep{korelc2010improved}, and 8-noded serendipity element in the axisymmetric one, see Section~S2.1 for details. Second, the two energy profiles exhibit differences in the vicinity of the imperfection (a region excluded from the profiles in Fig.~\ref{fig:SimSizeSum}), which further contributes to this discrepancy. Accordingly, as a way to visualize the relation between the two energy landscapes, the graph of the axisymmetric tube is shifted downward to align with that of the full tube at the smallest diameter $D=0.43$~mm, see the dashed curves in Fig.~\ref{fig:SimSizeEn}. This adjustment offers a clearer view on how the bulk energy trend would appear in the absence of morphological changes.

\begin{figure}
    \hspace*{0pt}
    \includegraphics[width=1\textwidth]{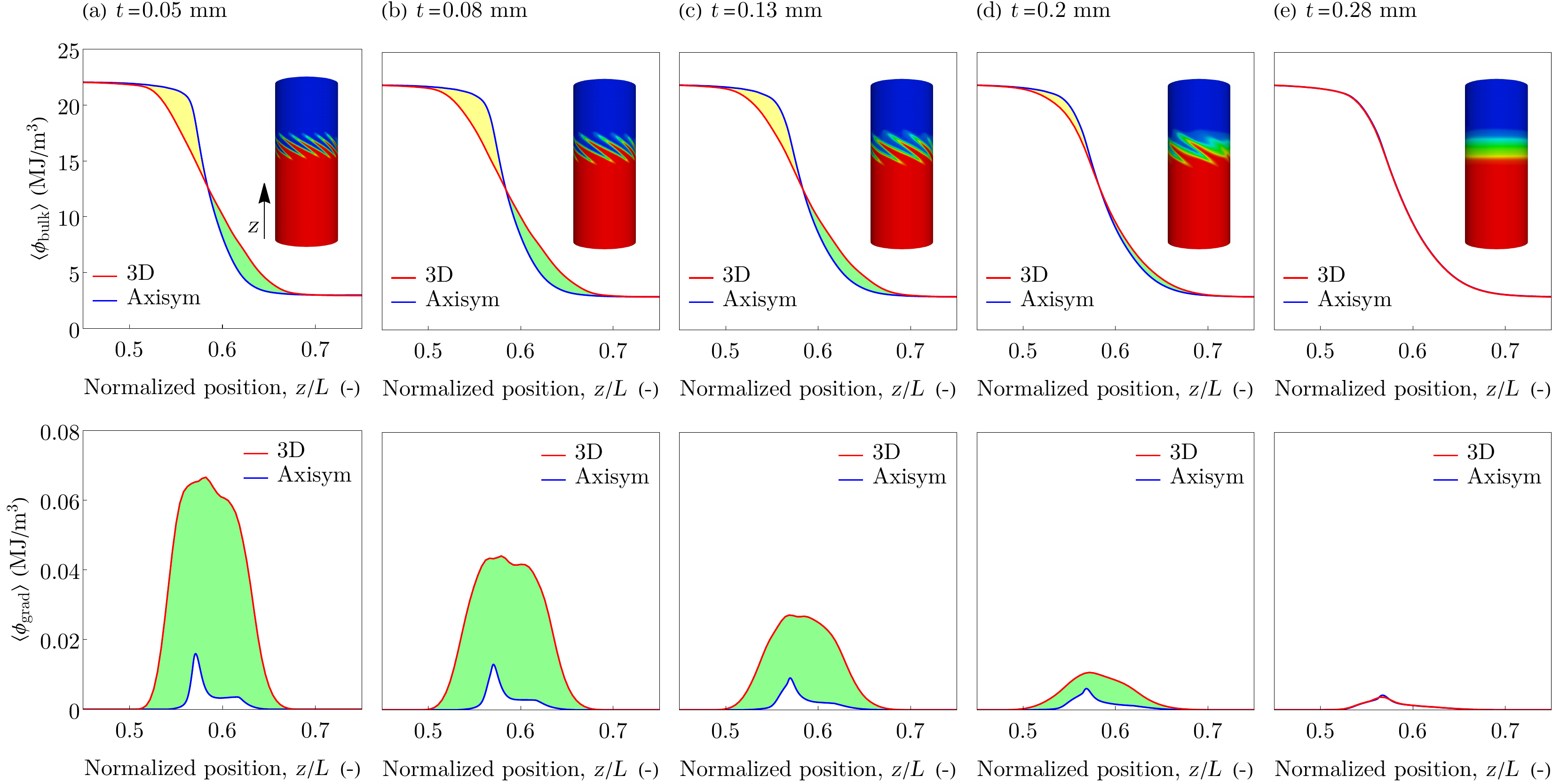}
    \caption{Profiles of the average bulk and gradient energy densities for varying tube thickness $t$ (recall that the tube diameter $D=2.5$~mm and tube length $L=6.25$~mm are fixed). In the graphs, greenish shading indicates the regions where the energy density of the full tube exceeds that of the axisymmetric reference, while yellowish shading indicates the opposite.}
    \label{fig:SimThickSum}
\end{figure}

\begin{figure}
    \hspace{0pt}
    \includegraphics[width=1\textwidth]{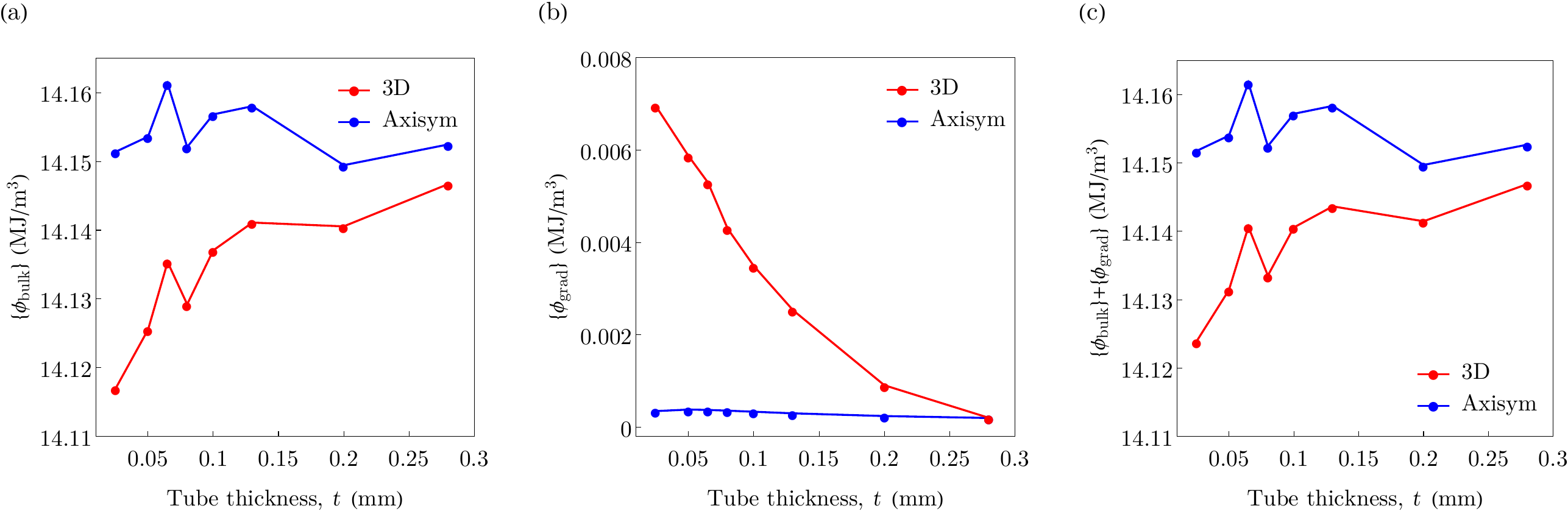}
    \caption{The overall average energy densities as a function of the tube thickness $t$: (a) the bulk energy density $\{ \phi_\text{bulk} \}$, (b) the gradient energy density $\{ \phi_\text{grad} \}$, and (c) the sum of the two contributions.}
    \label{fig:SimThickEn}
\end{figure}

We now examine the impact of the tube thickness $t$ on the average energy densities and front morphology, as illustrated in Figs.~\ref{fig:SimThickSum} and \ref{fig:SimThickEn}. The foremost observation is that as $t$ decreases, the transformation front adopts a morphology characterized by a higher number of fingers and sharper interfaces, in particular 7 fingers are observed in the thinnest (half-)tube, while a diffuse finger-less front is evolved in the thickest tube. Similar to the size effects discussed earlier, the transition towards a fingered morphology is accompanied by a deviation of the bulk energy density profile from its axisymmetric reference, with this deviation intensifying as the number of fingers increases. And obviously, a higher number of fingers corresponds to a larger gradient energy density.

The transition in front morphology with varying tube thickness $t$ can be attributed to the geometric mismatch (incompatibilities) between the austenite and martensite domains, which must be accommodated by elastic deformations. First, the radial component of the transformation strain (considering that the martensitic transformation is isochoric, the axial transformation strain is accompanied by compensating transformation strains in the transversal directions) induces an out-of-plane deflection in the martensite domain. This strain mismatch with the austenite domain is then accommodated by an out-of-plane bending within the austenite--martensite interface region. As $t$ increases, the bending stiffness of the tube increases, resulting in a higher energetic cost associated with this bending deformation. In response to this high cost, the energy minimization process prompts a more diffuse front to reduce the bending curvature. Indeed, this mechanism alone governs the morphological transitions in the axisymmetric tube configuration, where thicker tubes develop increasingly diffuse fronts and exhibit lower gradient energy densities, as shown in Fig.~\ref{fig:SimThickSum} (the front morphologies of the axisymmetric tubes are not provided here). Similar reasoning was proposed by \citet{he2009effects} to explain the variations in the geometry of the martensite helix for different tube thicknesses. 

Second, local adjustments required to accommodate the strain mismatches between the austenite and martensite fingers. Thicker tubes resist these local elastic deformations more strongly, rendering the formation of fingers energetically less favorable. In contrast, thinner tubes, being more compliant, can more easily accommodate a higher number of fingers and the associated local mismatches. Consequently, the observed (on-average) increase in the overall bulk energy density, $\{ \Phi_\text{bulk} \}$, as $t$ increases, and the corresponding decrease in its gradient energy competitor, as shown in Fig.~\ref{fig:SimThickEn}(a,b), can be directly linked to the increase in bending stiffness and the reduced vulnerability to accommodate the local strain mismatches.

\begin{figure}
    \centering
    \includegraphics[width=0.78\textwidth]{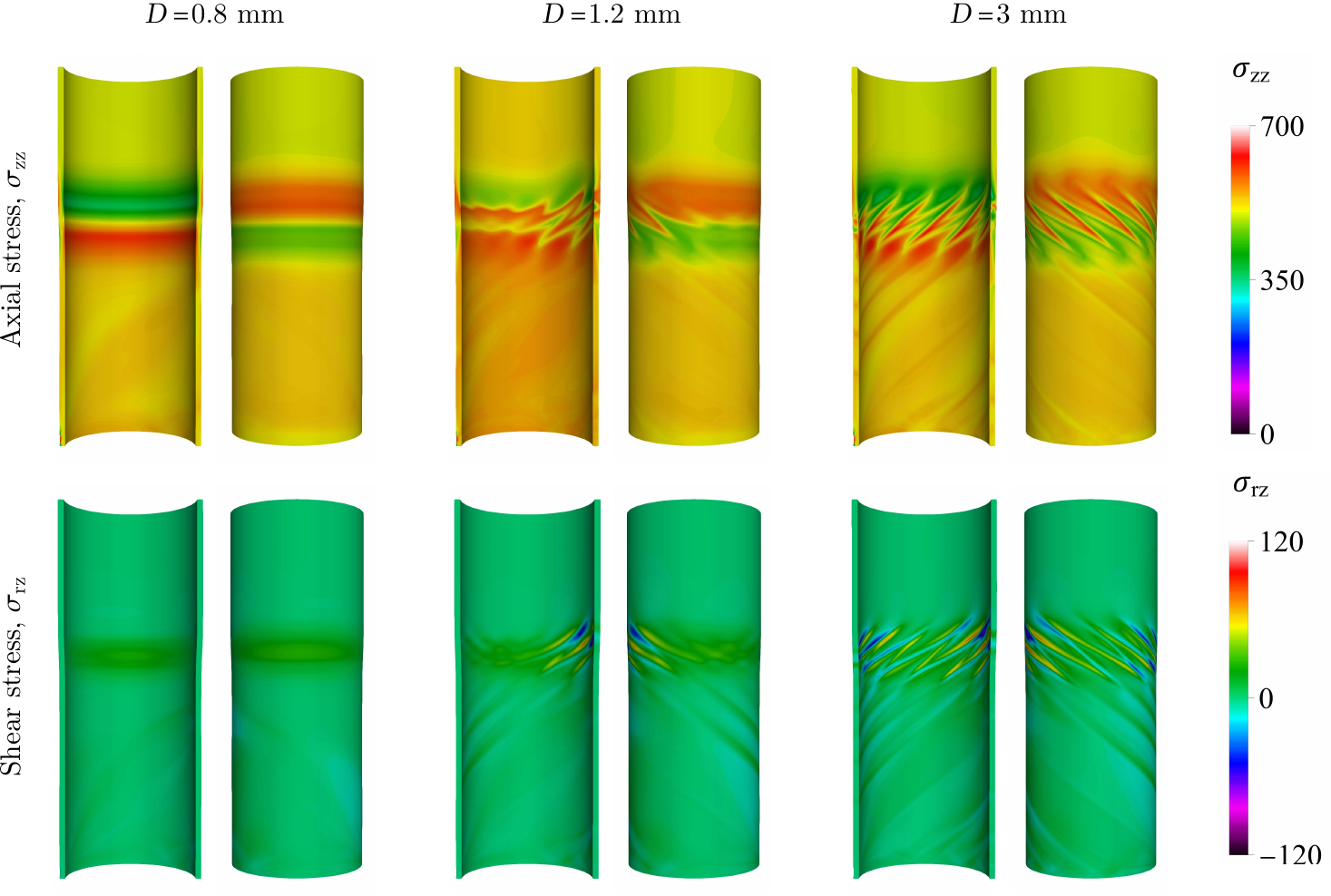}
    \caption{{Distribution of local stresses in tubes with different diameters $D$. The top and bottom rows show, respectively, the axial ($\sigma_\text{zz}$) and shear ($\sigma_\text{rz}$) components of the Cauchy stress tensor.}}
    \label{fig:stress}
\end{figure}

As a final remark of this investigation, we examine the local stress distributions in the tubes. Specifically, we analyze the tubes with diameters $D=3$~mm, $D=1.2$~mm, and $D=0.8$~mm, which develop markedly different transformation front morphologies (see Fig.~\ref{fig:SimSizeSum}). The corresponding axial and shear components of the Cauchy stress tensor, $\sigma_\text{zz}$ and $\sigma_\text{rz}$, are presented in Fig.~\ref{fig:stress}. Several interesting observations can be drawn from Fig.~\ref{fig:stress}. As expected, due to the localized nature of the phase transformation, all tubes exhibit significant stress inhomogeneities. The degree of stress inhomogeneity is closely tied to the morphology of the front: the more intricate and spatially refined the front, the more pronounced the resulting stress inhomogeneities. It can be seen that the axial stress displays a similar distribution in the inner wall of the tube, but with a reversed stress magnitude between the austenite and martensite regions. Additionally, the fine-scale stress inhomogeneities can differ between the inner and outer walls, which is particularly evident for the tube with $D=3$~mm. The shear stress $\sigma_\text{rz}$ reveals a qualitatively different behavior. In the tube with $D=0.8$~mm, where a diffuse finger-less front develops, the shear stress remains relatively uniform. In contrast, the larger tubes with fingered patterns exhibit pronounced shear stress concentrations, especially at the interfaces between austenite and martensite fingers. Notably, the peak shear stress in different tubes varies by up to 100~MPa. Similar trends can be also observed for other shear stress components.

It is important to emphasize that such strongly localized stress fields cannot be inferred from the global mechanical responses, which differ only slightly, mainly at stress events associated with transformation nucleation/saturation (plots of global mechanical responses are shown in Fig.~S3 in Supplementary Data). The pronounced local stress variations suggest that front morphology has a direct impact on fatigue behavior. This inference is in line with prior experiments on NiTi flat specimens subjected to cyclic subloop loading, which have shown that different front morphologies can lead to significantly different fatigue resistance \citep{zhang2018fatigue}.

\subsection{NiTi tubes under bending}\label{sec:analbend}
Unlike the preceding analysis, the focus here is not on a systematic energy-based investigation. Instead, by leveraging the insights gained from that analysis, the aim is to shed some light on the dimension-dependent transformation pattern and mechanical response in bending through interpretative comparisons of the simulation and experimental results. The shift in the analysis scope is dictated by two constraints. First, the unpredictability of nucleation events under bending deformation makes it difficult to establish a consistent transformation state at which the energy of different tubes can be directly compared. Second, the setup of the tube under uniaxial tension benefits from a simplified axisymmetric geometry, serving as a baseline for the energy and front morphology comparisons. However, no equivalent baseline is available for the tube under bending.

The material parameters are identical to those adopted in uniaxial tension simulations and remain consistent across all tube dimensions, see the details in Section~S2.2. To enrich our analysis, we not only report the results for the tubes tested in the experiment, namely the 2.5/0.28/8.93, 1.2/0.1/12, and 1.2/0.2/6.0 tubes, but also include the results for the remaining tubes that could not be tested due to the limitations discussed in Section~\ref{sec:expbendres}. Leveraging the symmetry of the tube under bending, only one-quarter of the entire tube geometry is simulated. For the tubes with $D=3$~mm and $D=2.5$~mm, the total length of the tube is taken as $2L=4.8 D$, where $L$ denotes the length of the simulated half-tube. For the remaining tubes, the total length is changed to $2L=7.5 D$. The length $2L$ is actually equal to the spacing between the two intermediate rollers in the bending setup, see Fig.~\ref{fig:setup}(e).

\begin{figure}
    \hspace*{0pt}
    \includegraphics[width=1\textwidth]{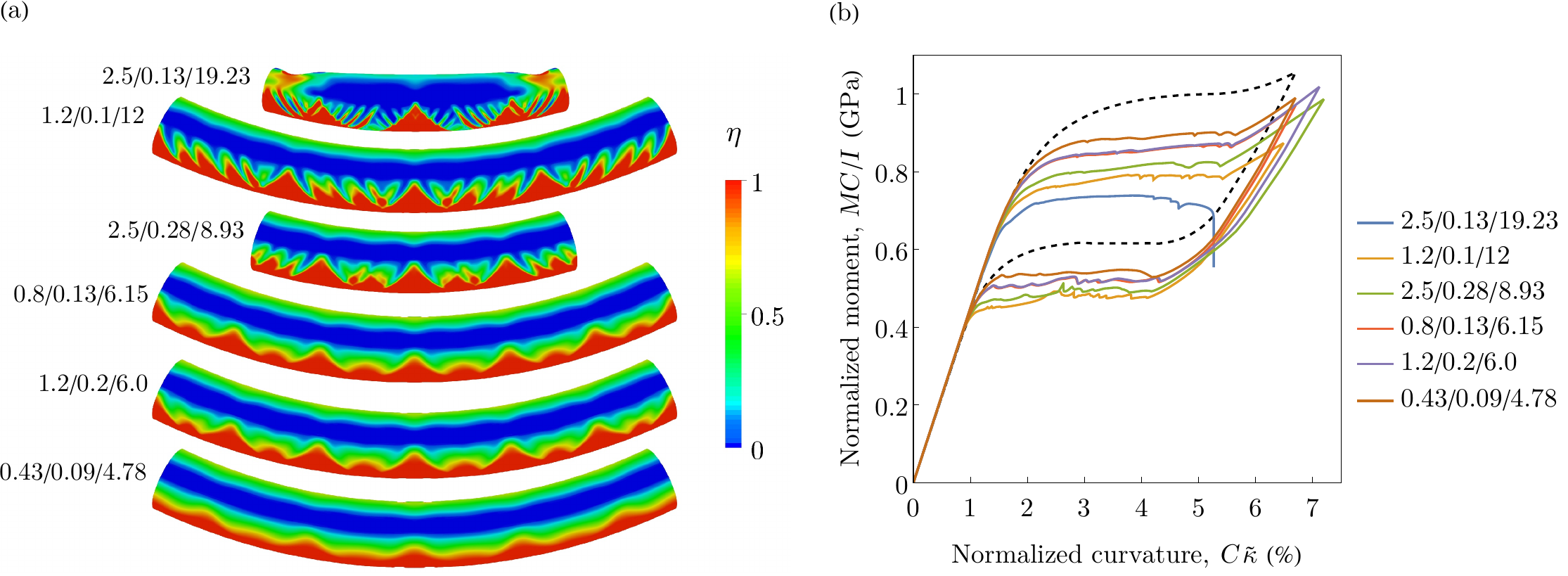}
    \caption{(a) Predicted transformation patterns in NiTi tubes of varying dimensions under bending, and (b) the corresponding normalized moment--curvature responses. {The dashed curve in panel (b) denotes the response obtained for NiTi rod (with a diameter of 0.43~mm).}}
    \label{fig:SimBend}
\end{figure}

The transformation patterns at the end of loading and the normalized moment--curvature responses are depicted in Fig.~\ref{fig:SimBend}. For a more lucid visualization of the patterns, the half (symmetric) tubes have been mirrored to appear as full tubes. The snapshots of the complete transformation evolutions are provided in the Supplementary Data, see Figs.~S4 and S5, and will be referenced to support the discussion and interpretation of the results. It becomes evident at first glance that all tubes share a prominent feature, that is, the formation of wedges on the tension side of the tube, while the transformation on the compression side remains fairly uniform. A closer inspection of the predicted transformation evolution in the tube geometries that were tested in the experiment, see Fig.~S4, reveals the following observations, some of which bear resemblance to experimental results:
\begin{enumerate}
    \item In the 2.5/0.28/8.93 and 1.2/0.1/12 tubes, the transformation evolves through the formation of three distinct wedges. As loading continues, the inter-wedge spacings are gradually filled by more randomly-oriented fingers. A subtle difference in the transformation pattern of these tubes lies in the degree of diffuseness. In the $D=2.5$~mm tube, the emerging wedges are enveloped within highly diffuse martensite regions. This pronounced diffuseness is also observable in the inter-wedge regions and persists in the final transformation pattern (see Fig.~\ref{fig:SimBend}(a)). This diffuseness is notably reduced in the $D=1.2$~mm tube with a larger diameter-to-thickness ratio of $D/t=12$ compared to $D/t=8.93$ in the $D=2.5$~mm tube.
    \item The transformation pattern in the 1.2/0.2/6.0 tube is somewhat different than the other two tubes. Specifically, five distinct closely-spaced wedges form and the inter-wedge spacings are subsequently filled with hump-like martensites, culminating in a recurring crown-shaped morphology. Note that the so-called crown-shaped morphology has been also observed in the experiment, see the related discussion in Section~\ref{sec:expbendresTP}, however, due to the limited deformation applied, it did not evolve into fully-transformed martensite.
\end{enumerate}

Among the four remaining tubes simulated, the 3.0/0.1/30 tube experiences early-stage local buckling on its compression side, and thereby, the corresponding results are excluded from further discussion. Buckling is also observed in the 2.5/0.13/19.23 tube, but it occurs after a considerable progression of the localized transformation. Indeed, of the presented transformation evolutions, the 2.5/0.13/19.23 tube exhibits the most intriguing one. In this tube, the transformation initiates with the nucleation of slim and sharp fingers (snapshot 1 in Fig.~S5), which later increase in number and arrange into an interesting radial configuration. Upon further loading, the radial arrangements of fingers transition into wedges, and at the same time, more fingers appear in the inter-wedge spacings. At this stage, the tube undergoes local buckling, and further bending cannot be simulated because of the lack of convergence of the global Newton method. The transformation evolution in the 0.8/0.13/6.15 tube mirrors that observed in the 1.2/0.2/6.0 tube, exhibiting similar characteristics throughout. Finally, the 0.43/0.09/4.78 tube, with the smallest diameter and the smallest diameter-to-thickness ratio, displays a highly diffuse transformation, with its detailed features closely aligning with those of the 0.8/0.13/6.15 and 1.2/0.2/6.0 tubes. Nevertheless, its final pattern appears more smooth and bears a resemblance to the transformation pattern reported for NiTi rod under bending \citep{watkins2018uniaxial}.

As elaborated in Section~\ref{sec:analten}, two primary factors govern the dimension-dependent transformation patterns: the size-dependent gradient energy contribution, which modulates the energy balance as the tube diameter varies, and the strain incompatibilities arising from the transformation strain within the martensite domain and the energetic cost of accommodating them in tubes of different thicknesses. The latter effect is clearly manifested within the range of tube geometries investigated in this study. This is evident when comparing tubes with identical diameters $D$ but differing $D/t$ ratios, such as the 2.5/0.28/8.93 versus 2.5/0.13/19.23 tubes, or the 1.2/0.2/6.0 versus 1.2/0.1/12 tubes. The former effect, although not as distinctly isolated, can still be inferred from the transformation patterns of tubes with comparable $D/t$ ratios but different $D$, e.g., the 2.5/0.28/8.93 and 1.2/0.2/6.0 tubes. By and large, the formation of wedges emerges as an inevitable characteristic of transformation evolution in tubes subjected to bending. Smaller $D/t$ ratios result in more diffuse and smoother wedges, while simultaneously reducing the susceptibility of inter-wedge spacings to being filled with randomly oriented martensite fingers. This results in a wedge-to-flat transition of the transformation pattern as $D/t$ increases, a trend clearly illustrated in Fig.~\ref{fig:SimBend}(a).

We conclude this discussion by addressing the moment--curvature responses of the tubes, as shown and compared in Fig.~\ref{fig:SimBend}(b). Consistent with the experimental results, the normalized bending moment $MC/I$ and the normalized curvature $C \tilde{\kappa}$ are used to describe the mechanical behavior of the tubes. The only distinction lies in the definition of the curvature $\tilde{\kappa}$, calculated as $\tilde{\kappa} = (\varphi_1 - \varphi_2)/2L$, where the normalization is based on the end-to-end length $L$ rather than $L_\text{e}$, the length of AOI. The results reveal trends qualitatively consistent with the experimental findings. Specifically, smaller $D/t$ ratios correspond to higher elastic limits and thus higher moment-plateau levels, such that the response obtained for a rod geometry (see the dashed curve) represents an upper bound of the moment plateau. The two 1.2/0.1/6.0 and 0.8/0.13/6.15 tubes, with nearly the same $D/t$ ratio, exhibit almost identical mechanical responses, matching with their transformation patterns. Moreover, the responses highlight a geometric influence on the moment hysteresis loop area as well; smaller $D/t$ ratios result in larger hysteresis areas, in line with the experimental observations. Finally, it is worth noting that nearly all the responses display a mild hardening behavior. Instances of softening can also be observed, which correspond to the nucleation of martensite fingers within the inter-wedge spacings.

\section{Concluding remarks}
Distinct transformation patterns emerge in superelastic NiTi tubes under mechanical loading. Helical domains, branched front morphologies, and criss-cross patterns have been observed in tension experiments. Under large-rotation bending, fingered martensite bands nucleate, grow, and merge to form wedges. These complex patterns have been consistently reported in prior experiments in the literature, and are generally known to be dimension-dependent. In this study, we systematically address the effect of (outer) tube diameter ($D$) and wall-thickness ($t$) on the transformation pattern and front morphology in quasi-static tension and large-rotation bending experiments through real-time stereo-DIC measurements. The experimental observations are corroborated with finite-element modeling and the underlying mechanisms are examined. The following conclusions can be drawn:
\begin{itemize}
  \item In tension, no systematic dependence of the global mechanical responses on the tube dimensions in the $0.43~\text{mm} \leq D \leq 3~\text{mm}$ and $4.8 \leq D/t \leq 30$ ranges has been observed. All tested tubes exhibit plateau-type stress--strain superelasticity with a large stress hysteresis, despite showing various transformation patterns and front morphologies. 
  \item The slimness and number of martensite bands (fingers) that nucleate at the onset of the stress-plateau, along with the fineness of the austenite--martensite front morphology (formed by the merger of the fingers) within the stress-plateau region vary systematically with $D$ and $D/t$. As $D$ and $D/t$ increase, i.e., in larger and thinner tubes, an increasing number of slimmer fingers tend to nucleate, resulting in a more finely-fingered austenite--martensite front. 
  \item In-situ observations, as also supported by modeling, reveal a gradual transition in the front morphology with increasing $t$: evolving from a finely-fingered front to a coarsely-fingered front characterized by highly-diffuse fingers, and ultimately turning into a flat finger-less front. The $t$-dependence of the equilibrium front morphology also depends on $D$, i.e., there is an interplay between $D$ and $t$ that dictates the equilibrium front morphology. 
  \item Modeling reveals that the underlying mechanism behind the front morphology transition for varying $D$ lies in the role of the gradient (interfacial) energy, which becomes increasingly dominant and governing for smaller tube sizes. At the same time, as the tube becomes thicker, the energetic cost associated with accommodating strain incompatibilities within the austenite and martensite fingers becomes more significant, leading to a shift in the energy balance that favors the development of fewer fingers. Interestingly, the local axial stress exhibits pronounced localization within the front, irrespective of its morphology. In contrast, the shear stress, while being nearly uniform in finger-less morphology, is highly concentrated at the austenite--martensite fingers. The distinct local stress distribution arising from the variation in the strain localization behavior can strongly impact fatigue performance.
  \item In contrast to the global responses in tension, the moment--curvature responses of the NiTi tubes under bending show a relatively considerable dependence to $D/t$. As $D/t$ decreases, the elastic strain that the tube can sustain prior to nucleation of the first martensite band increases, leading to a shift in the (upper and lower) moment plateaus. The overall moment--curvature response exhibits a stronger hardening as $D/t$ decreases, however, there are strain spans in which the moment shows a decreasing trend. This size-dependent behavior has been clearly reproduced in the simulations.
  \item Slim and sharp martensite fingers nucleate, grow, and merge to form wedges in the tensile region of larger tubes under bending. In smaller tubes, the wedges emerge from much more diffuse martensite fingers. At the later stage of transformation, the accommodation of large bending deformation is manifested differently in different tube geometries. Specifically, the inter-wedge regions are either filled with martensite fingers or remain weakly transformed leading to a crown-shaped morphology.
  \item Simulation results support the experimental findings on the evolution of the transformation pattern under pure bending. The formation of the diffuse and connected wedges in smaller tubes with lower $D/t$ ratios can be attributed to the combined effects of a high gradient energy contribution and a large energetic cost of accommodating strain incompatibilities, analogous to the mechanisms observed under uniaxial tension. However, due to the inherent complexities of the bending setup, establishing a simple and consistent baseline for bending {(like the axisymmetric tube in tension)} is not possible. It has been observed that for the smallest $D/t$, the transformation pattern is similar to that observed in NiTi rod under bending.
\end{itemize}

The results presented in this study carry important implications for various applications of NiTi tubes. Elastocaloric heat pumps that utilize bending deformation are gaining attention due to their lower mechanical load requirement. Our findings suggest that thinner tubes may offer advantages in such cooling technologies, not only due to higher surface-to-volume ratio, but also due to the lower moment required to initiate phase transformation, larger transformed volume and enhanced latent heat, and reduced hysteresis loop area. In contrast, thicker tubes exhibit higher bending moments during transformation, which may adversely affect their fatigue performance.

The observations and conclusions drawn in this study may have relevance beyond superelastic NiTi alloys, particularly for material systems that exhibit PT-like inelastic mechanism. Deformation twinning is one such mechanism, which serves as a dominant mode of deformation in many metals and alloys, such as magnesium. The interaction between deformation twinning and plasticity gives rise to complex microstructural patterns at the grain level. Experiments on magnesium under bending have revealed the intriguing presence of needle-like twins \citep{mcclelland2015geometrically,yoshida2024analysis}. The needles, to some qualitative extent, resemble the transformation pattern in NiTi tubes, and thus their formation and evolution could potentially be explained through an energy-minimization methodology.

\paragraph{\textbf{Acknowledgments}}
We gratefully acknowledge the financial support, provided by Pasargad Institute for Advanced Innovative Solutions (PIAIS) under grant numbers 10104 and 10159. Aslan Ahadi would like to thank the Alexander von Humboldt foundation for sponsoring an experienced-researcher fellowship. Gunther Eggeler and Jan Frenzel received funding from the Deutsche Forschungsgemeinschaft (DFG) under the project 498172553. Mohsen Rezaee-Hajidehi acknowledges the financial support by the National Science Centre (NCN) in Poland through the Grant No.\ 2021/43/D/ST8/02555. We would like to thank Reza Malekzadeh for his assistance with manufacturing bending fixtures and Javad Charkhchian for helping with preliminary tensile experiments. For the purpose of Open Access, the authors have applied a CC-BY public copyright license to any Author Accepted Manuscript (AAM) version arising from this submission.

\bibliographystyle{elsarticle-harv}

%\bibliographystyle{unsrtnat}

%\bibliographystyle{elsarticle-num}
%\biboptions{sort&compress}
%\biboptions{sort}

\bibliography{bibliography}

\clearpage


\section*{Supplementary material}

# Supplementary Data for:
# Size-dependent transformation patterns in NiTi tubes under tension and bending: stereo digital image correlation experiments and modeling

Aslan Ahadi[a,b,*], Elham Sarvari[a], Jan Frenzel[b], Gunther Eggeler[b], Stanisław Stupkiewicz[c], Mohsen Rezaee-Hajidehi[c,**]

[a]*Pasargad Institute for Advanced Innovative Solutions (PIAIS),*
*Tehran, Iran.*
[b]*Institute for Materials, Ruhr University Bochum,*
*Universitätsstr. 150, 44801 Bochum, Germany*
[c]*Institute of Fundamental Technological Research (IPPT), Polish Academy of Sciences,*
*Pawińskiego 5B, 02-106 Warsaw, Poland.*

**Abstract**

This document contains the supplementary data related to the NiTi tubes used in the tension and bending experiments, along with the model parameters used in the simulations and additional simulation results.

## S1. Experiment

*S1.1. DSC measurements and grain size*

The DSC thermograms for all tubes investigated in the experiment and the histograms of grain size distribution of four representative tubes are shown in Figs. S1 and S2, respectively.

[*]Corresponding author of the experimental part.
[**]Corresponding author of the modeling part.
*Email addresses:* `aslan.ahadi@piais.ir,aslan.ahadi@rub.de` (Aslan Ahadi), `elham.sarvari@piais.ir` (Elham Sarvari), `jan.a.frenzel@rub.de` (Jan Frenzel), `gunther.eggeler@rub.de` (Gunther Eggeler), `sstupkie@ippt.pan.pl` (Stanisław Stupkiewicz), `mrezaee@ippt.pan.pl` (Mohsen Rezaee-Hajidehi)

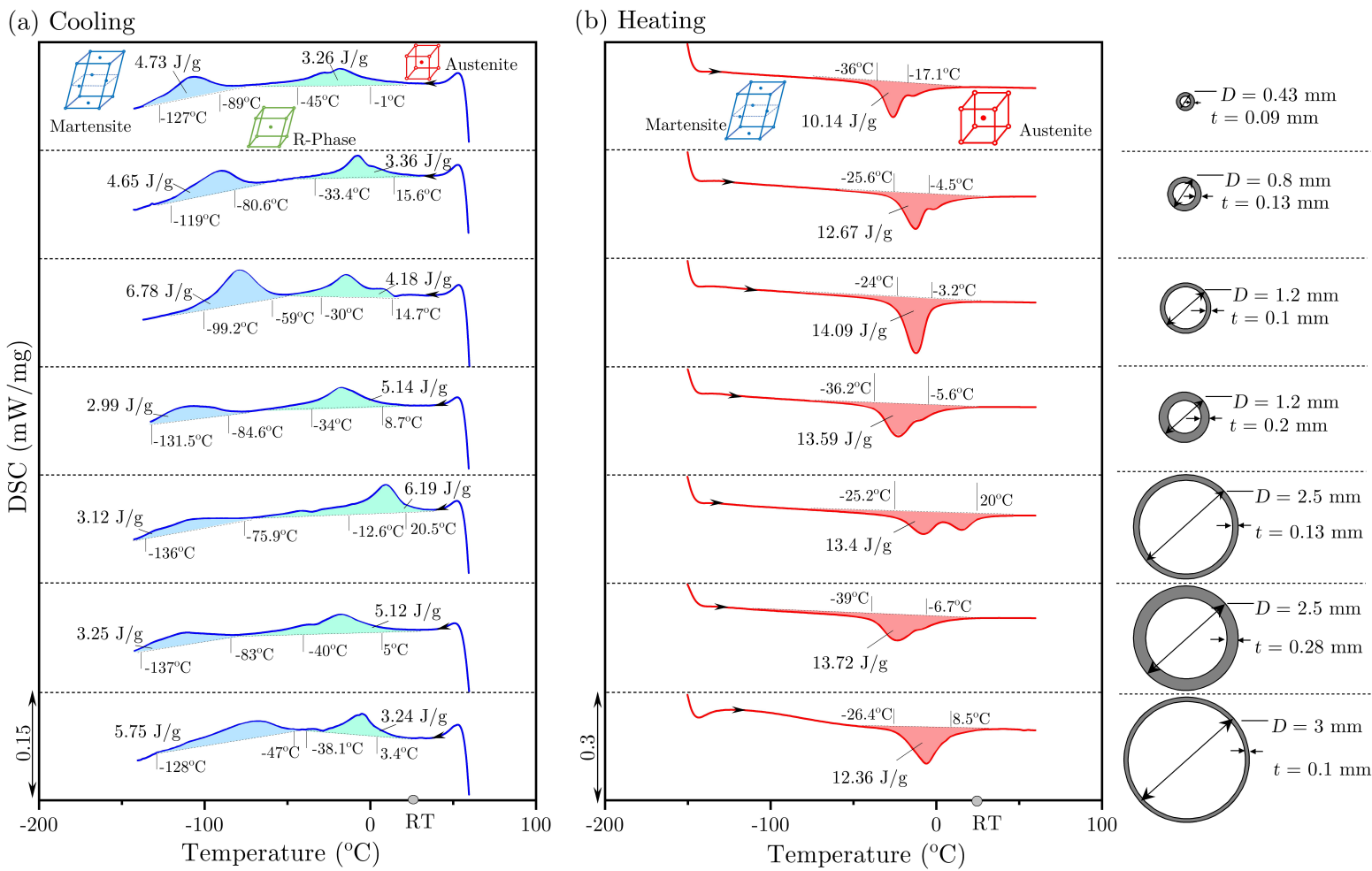

Figure S1: DSC thermograms during cooling (a) and heating (b) of the NiTi tubes with different $D$ and $D/t$ ratios. RT stands for room temperature. Note that for all tubes RT is higher than $A_f$ suggesting superelastic responses at RT.

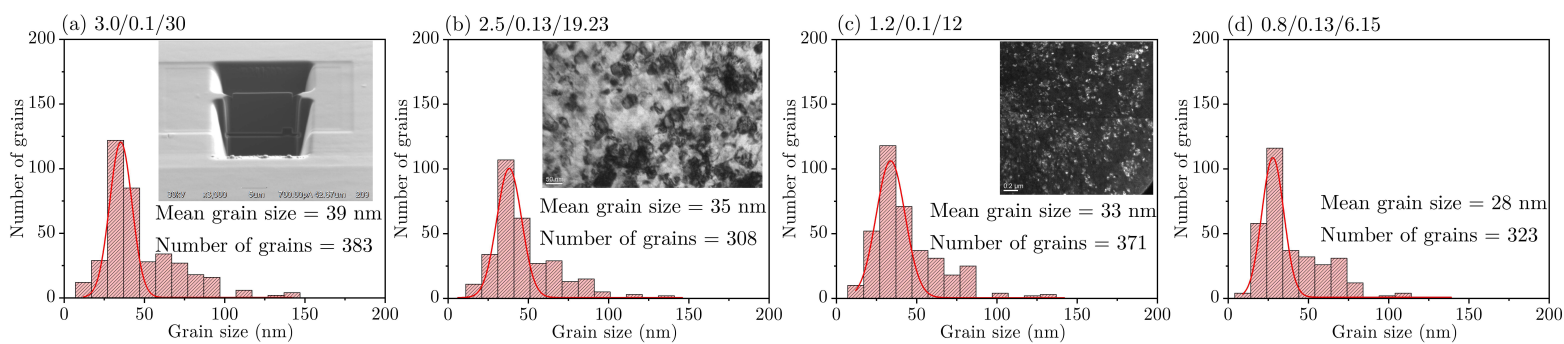

Figure S2: Histograms of grain size distribution of (a) 3.0/0.1/30, (b) 2.5/0.13/19.23, (c) 1.2/0.1/12, and (d) 0.8/0.13/6.15 tubes. The inset in (a) is a thin slab for TEM observations prepared by FIB. Inset in (b) is a typical bright-field image, and in (c) is a low-mag dark-field image that is used for measuring grain size distributions.

2

*S1.2. Specifications of the NiTi tubes*

The specimen and gauge area dimensions, speckling parameters, speckle pattern features, DIC parameters, and correlation analysis details of the NiTi tubes tested in the quasi-static tension and large-rotation bending experiments are provided in Tables S1 and S2, respectively.

Table S1: First row: Specimen dimensions used in quasi-static tension experiments. Second row: Speckling parameters and speckled pattern features. Third row: DIC parameters

| parameters | $D = 3.0$ mm $D/t = 30$ | | $D = 2.5$ mm $D/t = 19.23$ | | $D = 1.2$ mm $D/t = 12$ | | $D = 2.5$ mm $D/t = 8.93$ | | $D = 0.8$ mm $D/t = 6.15$ | | $D = 1.2$ mm $D/t = 6$ | | $D = 0.43$ mm $D/t = 4.78$ | |
|---|---|---|---|---|---|---|---|---|---|---|---|---|---|---|
| | $T_1$ | $T_2$ | $T_1$ | $T_2$ | $T_1$ | $T_2$ | $T_1$ | $T_2$ | $T_1$ | $T_2$ | $T_1$ | $T_2$ | $T_1$ | $T_2$ |
| $L_f$ (mm) | 11 | 12.5 | 10.25 | 10.12 | 5.65 | 5.1 | 11.5 | 10.04 | 4.5 | 4.3 | 6 | 5.2 | 4.6 | 5 |
| $L_f/D$ | 3.67 | 4.17 | 4.1 | 4.05 | 4.7 | 4.25 | 4.6 | 4.02 | 5.63 | 5.38 | 5 | 4.33 | 10.7 | 11.63 |
| $L_e$ (mm) | 10.2 | 10.88 | 9.88 | 9.56 | 5.25 | 4.7 | 11.25 | 9.6 | 3.88 | 3.7 | 5.68 | 4.95 | 4.3 | 4.76 |
| $L_e/D$ | 3.4 | 3.63 | 3.95 | 3.82 | 4.37 | 3.92 | 4.5 | 3.84 | 4.85 | 4.63 | 4.7 | 4.13 | 10 | 11.07 |
| $\dot{\delta}/L_f$ ($10^{-5}$ s$^{-1}$) | 7.5 | 6.6 | 8.1 | 8.2 | 14.7 | 16.3 | 7.2 | 8.3 | 9.2 | 9.7 | 14.35 | 16 | 18 | 16.5 |
| Pressure (psi) | 45 | 45 | 50 | 50 | 60 | 60 | 50 | 50 | 60 | 60 | 60 | 60 | 75 | 75 |
| Distance (cm) | 30 | 30 | 30 | 30 | 30 | 30 | 30 | 30 | 30 | 30 | 30 | 30 | 35 | 35 |
| Nozzle # | 2 | 2 | 2 | 2 | 2 | 2 | 2 | 2 | 1 | 1 | 2 | 2 | 1 | 1 |
| $\rho_s$(%) | 35.1 | 40.3 | 40.1 | 38.7 | 48.6 | 49.7 | 48.6 | 58.1 | 54 | 41 | 55.3 | 47.1 | 44 | 38 |
| Gray value | 204 | 221 | 213 | 201 | 199 | 205 | 205 | 193 | 204 | 214 | 197 | 203 | 219 | 209 |
| $\overline{\nabla G}$ | 136 | 118 | 107.1 | 114 | 98.7 | 103 | 106.8 | 89 | 94 | 121 | 91 | 97 | 116 | 134 |
| Magnification (X) | 0.5 | 0.5 | 0.53 | 0.53 | 1 | 1 | 0.53 | 0.53 | 1.5 | 1.5 | 1 | 1 | 1.5, 2 | 1.5, 2 |
| Exposure time (ms) | 109 | 95 | 165 | 180 | 117 | 130 | 145 | 165 | 183 | 203 | 132 | 133 | 115 | 104 |
| Pix per mm (pix) | 254 | 254 | 289 | 289 | 547 | 547 | 289 | 289 | 717 | 719 | 545 | 545 | 719 | 719 |
| Anal. grid dim. (pix) | 665×2591 | 668×2764 | 650×2834 | 612×2744 | 531×2825 | 541×2525 | 625×3540 | 631×2688 | 469×2658 | 488×2763 | 539×3076 | 526×2633 | 225×3091 | 211×3422 |
| Subset size (pix) | 59×59 | 53×53 | 57×57 | 93×93 | 99×99 | 55×55 | 73×73 | 85×85 | 101×101 | 109×109 | 107×107 | 105×105 | 83×83 | 93×93 |

Table S2: First block: specimen dimensions used in large-rotation bending experiments. Second block: speckling parameters and speckled pattern features. Third block: DIC parameters

| parameters | $D = 1.2$ mm $D/t = 12$ | $D = 2.5$ mm $D/t = 8.93$ | $D = 1.2$ mm $D/t = 6$ |
|---|---|---|---|
| $L_\text{f}$ (mm) | 7.2 | 11.5 | 7.2 |
| $L_\text{f}/D$ | 6 | 4.6 | 6 |
| $L_\text{e}$ (mm) | 4.6 | 10 | 4.85 |
| $L_\text{e}/D$ | 3.83 | 4 | 4.04 |
| $\dot{\delta}$ (mm/s) | 0.0017 | 0.0017 | 0.0017 |
| Pressure (psi) | 60 | 55 | 60 |
| Distance (cm) | 30 | 30 | 30 |
| Nozzle # | 2 | 2 | 2 |
| $\rho_\text{s}$(%) | 53 | 44.3 | 50.4 |
| Gray value | 213 | 217 | 207 |
| $\overline{\nabla G}$ | 126 | 149 | 115 |
| Magnification (X) | 1 | 0.5 | 1 |
| Exposure time (ms) | 28 | 156 | 35 |
| Pix per mm (pix) | $\sim$545 | $\sim$255 | $\sim$545 |
| Anal. grid dim. (pix) | 556$\times$3455 | 569$\times$2810 | 523$\times$3482 |
| Subset size (pix) | 57$\times$57 | 65$\times$65 | 59$\times$59 |

## S2. Modeling

*S2.1. Finite-element implementation of the model*

The model presented in Section 5.1 is first made thermomechanically coupled and is subsequently implemented within the finite-element method. The thermomechanical coupling process involves extending the chemical energy of transformation, see Eq. (5) in the main paper, to account for the Clausius–Clapeyron relationship, and introducing an internal heat source that comprises the latent heat of transformation and the heat produced by mechanical dissipation, see the detailed formulation in Rezaee-Hajidehi et al. (2020). The finite-element implementation requires various treatments that are crucial for a successful and robust performance. Foremost among these is the micromorphic regularization that enables a straightforward implementation of the gradient-enhanced model Forest (2009), see also Mazière and Forest (2015); Rezaee Hajidehi and Stupkiewicz (2018). The micromorphic regularization is accomplished by introducing a new global variable, denoted by $\breve{\eta}$, and coupling it to the volume fraction $\eta$ via the penalty method. This is done by adding an energy penalization term into the Helmholtz free energy function $\phi$, cf. Eq. (3), i.e.,

$$\phi = \phi_{\text{bulk}} + \phi_{\text{grad}} + \phi_{\text{pen}}, \qquad \phi_{\text{pen}} = 1/2\chi(\eta - \breve{\eta})^2, \tag{S1}$$

with $\chi$ as the penalization parameter. Upon this modification, the gradient energy $\phi_{\text{grad}}$, see Eq. (8), can be reformulated in terms of $\nabla \breve{\eta}$ rather than $\nabla \eta$. The variable $\eta$ thus becomes a local variable, and its governing equation (along with that of the transformation strain $\bar{e}^{\text{t}}$) can be solved point-wise at the integration-point level, thereby simplifying the numerical treatment. Additionally, to handle the non-differentiable rate-independent dissipation and the inequality constraint $0 \leq \eta \leq 1$, the augmented Lagrangian method is employed Stupkiewicz and Petryk (2013). The resulting finite-element model adopts a global–local structure, necessitating a nested iterative–subiterative solution scheme in which the Newton method is utilized to solve the global problem and the local sub-problems.

The model possesses three global unknown fields: the displacement, the micromorphic variable and the temperature. An important consideration is the choice of elements in spatial discretization, such that a reliable prediction is achieved while maintaining a reasonable mesh density and computational cost. In the present context, a commonly adopted strategy involves the use of 20-noded serendipity elements with a reduced integration rule for the displacement, paired with linear 8-noded elements for the micromorphic variable and temperature Xiao and Jiang (2020); Rezaee-Hajidehi and Stupkiewicz (2021); Tsimpoukis et al. (2023). However, our preliminary simulations indicate that the best performance is obtained when 8-noded TSCG12 (12-mode enhanced-assumed strain) elements with a special 9-point integration rule are adopted for the displacement Korelc et al. (2010). Accordingly, the TSCG12 element is selected for 3D simulations. For the 2D axisymmetric simulations, 8-noded serendipity elements are used for the displacement field, and linear elements are used for the micromorphic and temperature.

The AceGen system Korelc and Wriggers (2016) has been employed to develop the finite-element computer code. The automatic differentiation (AD) tool of AceGen along with its AD exception capability greatly streamlines the computer implementation process and enhances significantly the efficiency of the resulting finite-element model.

*S2.2. Material parameters*

We have adopted an identical set of material parameters for all the quasi-static tension and bending finite-element simulations. A more elaborate approach would involve calibrating the parameters individually for each tested tube. However, this would introduce additional sources of variability and could obscure a clear interpretation of the analyses outcome. It is also noteworthy that the parameters governing tension–compression asymmetry and transverse isotropy (with the former having a direct impact on the bending response) are defined according to prior modeling experiences on tubes under combined loading conditions (Rezaee-Hajidehi and Stupkiewicz, 2021). A more precise calibration of these parameters would require inputs from experiments on tubes under uniaxial compression and torsion, which fall beyond the scope of our experimental work. As a consequence, while it is plausible to conjecture that the transformation patterns and mechanical responses observed in the experiment are influenced, to some extent, by intrinsic differences in the tested tubes, such differences, potentially representable by varying material parameters, do not form the main concern in our numerical analyses and thus are not accounted for.

The material parameters are listed in Table S3. For a full description of the model formulation and the role of the material parameters in the constitutive equations, the reader is referred to Rezaee-Hajidehi (2024). The parameters characterizing the trilinear material response are calibrated against the experimental response of the 2.5/0.13/19.23 tube, the T1 experiment to be specific, see Fig. 2 in the main paper. The calibration process involves adjusting the material parameters such that the so-called Maxwell construction of the trilinear material response aligns with the stress-plateau of the experimental curve during both loading and unloading. Additionally, the tension–compression asymmetry and transverse isotropy features are accounted for. Concerning the gradient energy parameter $G$, a reasonable value is first assigned to the theoretical interface thickness $\lambda$, specifically $\lambda = 40$ $\mu$m. The parameter $G$ is then calculated using the relation $G = -H\lambda^2/\pi^2$ (with $H < 0$ denoting the softening modulus in tension), see also the discussion following Eq. (8). It is important to remark that in 1D/2D settings, where regularization is governed exclusively by gradient effects, the parameter $\lambda$ corresponds to the actual thickness of a (planar) austenite–martensite interface. In 3D problems, additional regularization effects emerge from the 3D deformation state. As a result, the actual interface thickness becomes larger than the theoretical value $\lambda$, see a detailed discussion in Rezaee-Hajidehi et al. (2020).

Table S3: Parameters adopted in the finite-element computations.

| | Parameter | Unit | Value |
|---|---|---|---|
| $\Delta s^*$ | Specific entropy difference | MPa/K | 0.24 |
| $\kappa$ | Bulk modulus | GPa | 150 |
| $\mu_\text{a}$ | Shear modulus for austenite | GPa | 15.5 |
| $\mu_\text{m}$ | Shear modulus for martensite | GPa | 8.6 |
| $T_\text{t}$ | Transformation equilibrium temperature | K | 218 |
| $f_\text{c}$ | Dissipation parameter | MPa | 2.5 |
| $f_\text{r}$ | Dissipation parameter | MPa | 2.5 |
| $\epsilon_\text{T}$ | Maximum transformation strain in tension | [-] | 3.8% |
| $H_\text{T}$ | Softening modulus in tension | MPa | -5 |
| $H_\text{C}$ | Hardening modulus in compression | MPa | 8 |
| $\alpha$ | Tension–compression asymmetry ratio | [-] | 1.4 |
| $\beta$ | Transverse isotropy parameter$^\star$ | [-] | 1.0 |
| $G$ | Gradient energy parameter | Pa mm$^2$ | 810 |
| $\chi$ | Micromorphic penalty parameter | MPa | 500 |
| $K$ | Heat conductivity coefficient | W/(m K) | 18 |
| $\varrho_0 c$ | Specific heat capacity | MJ/(m$^3$K) | 2.86 |

$^\star$ The axis of transverse isotropy is aligned with the longitudinal axis of the tube.

*S2.3. NiTi tubes under quasi-static tension*

The global mechanical responses of three representative simulated tubes with different diameters $D$ are shown in Fig. S3. The arrow on the plot indicates the point at which the spatial stress inhomogeneities are compared in Fig. 12. While some small discrepancies are observable, particularly in the region where phase transformation interacts with the bottom end of the tube and eventually saturates (see the inset), the mechanical responses are remarkably similar. However, as shown and discussed in the paper, the corresponding front morphologies and the resulting spatial stress distributions differ significantly.

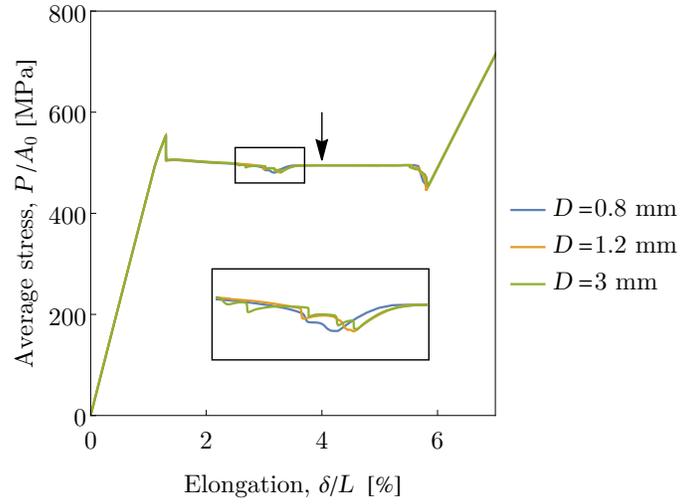

Figure S3: The global mechanical responses of simulated tubes with different diameters $D$.

## S2.4. NiTi tubes under bending

Snapshots of the predicted transformation pattern evolutions in the tubes under bending are shown in Fig. S4 and Fig. S5.

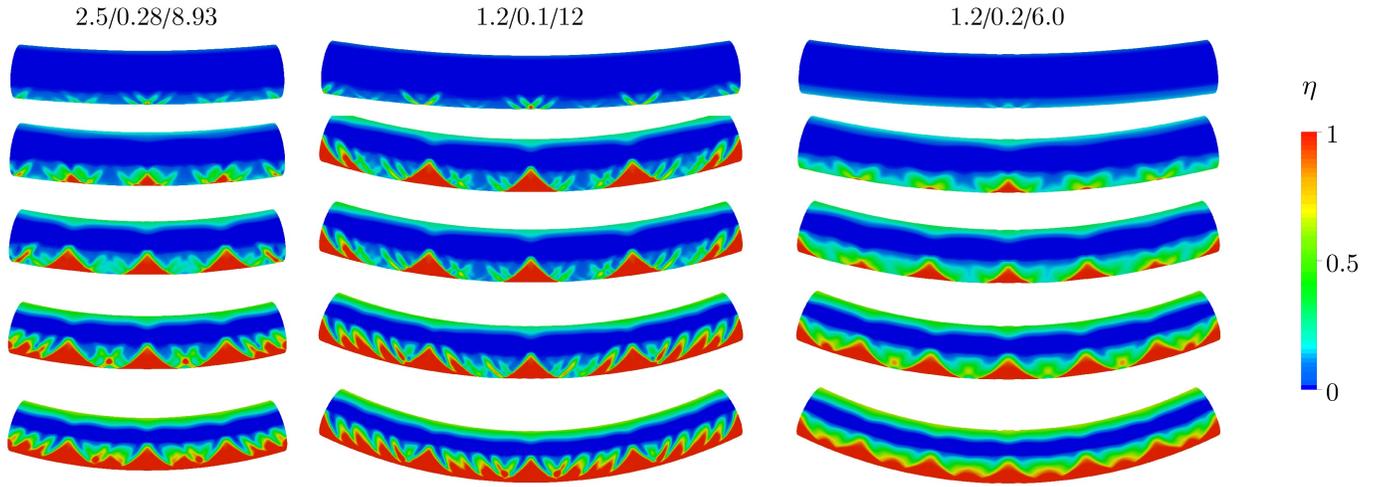

Figure S4: Snapshots illustrating the evolution of transformation patterns in NiTi tubes of varying dimensions under bending. These tube dimensions have been examined in the experiment.

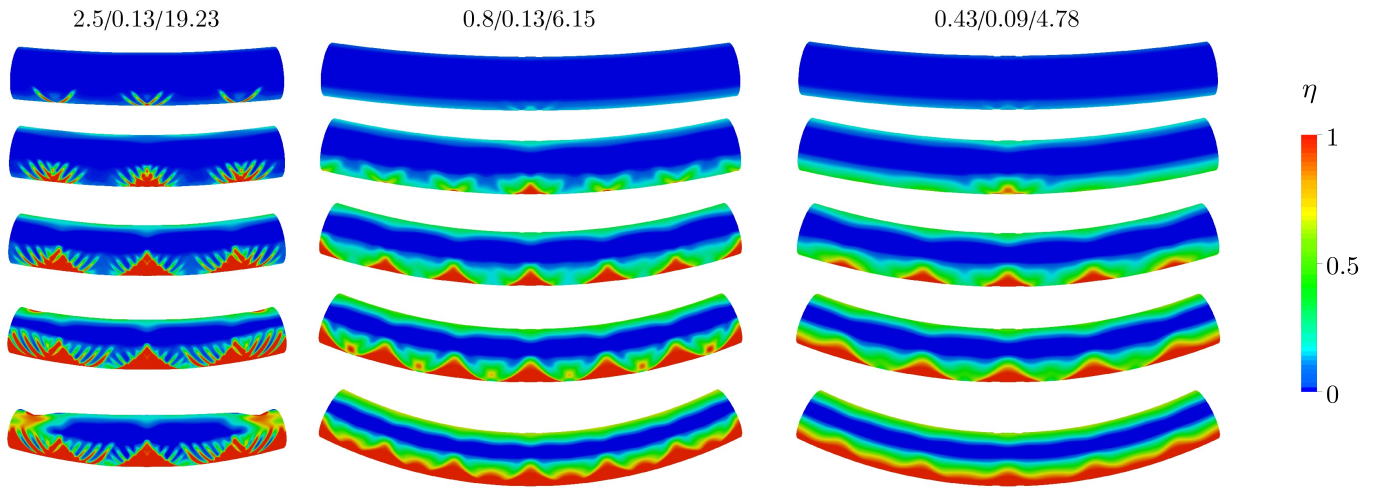

Figure S5: Snapshots illustrating the evolution of transformation patterns in NiTi tubes of varying dimensions under bending. These tube dimensions have not been examined in the experiment, see the related discussion in Section 3.4 in the main paper.



\end{document}